\documentclass{ws-jai}
\usepackage[flushleft]{threeparttable}
\usepackage[font=footnotesize]{subcaption}

\usepackage{float}
\usepackage{booktabs}

\begin{document}

\catchline{}{}{}{}{} 

\markboth{Author's Name}{Paper Title}

\title{Autofocusing Optimal Search Algorithm for a Telescope System\\
}

\author{Islam Helmy$^{1}$,  Alaa Hamdy$^{2}$, Doaa Eid$^{1}$,  and Ahmed Shokry$^{1}$ }

\address{
$^{1}$Department of Astronomy, National Research Institute of Astronomy and Geophysics, Helwan, Egypt\\
$^{2}$Department of Electronics Communications and Computer Engineering, Faculty of Engineering, Helwan University,
Helwan, Egypt\\
}

\maketitle

\corres{$^{2}$Corresponding author.}

\begin{history}
\received{(to be inserted by publisher)};
\revised{(to be inserted by publisher)};
\accepted{(to be inserted by publisher)};
\end{history}

\begin{abstract}
Focus accuracy affects the quality of the astronomical observations. Auto-focusing is necessary for imaging systems designed for astronomical observations. The automatic focus system searches for the best focus position by using a proposed search algorithm. The search algorithm uses the image's focus levels as its objective function in the search process. This paper aims to study the performance of several search algorithms to select a suitable one. The proper search algorithm will be used to develop an automatic focus system for Kottamia Astronomical Observatory (KAO). The optimal search algorithm is selected by applying several search algorithms into five sequences of star-clusters observations. Then, their performance is evaluated based on two criteria, which are accuracy and number of steps. The experimental results show that the Binary search is the optimal search algorithm.
\end{abstract}

\keywords{Autofocus; telescope system; Binary search; astronomical.}

\section{Introduction}
\label{intro}
High-quality astronomical observations are extremely important for astronomical research. The quality and accuracy of astronomical observations are affected by various factors such as seeing conditions, light pollution, and CCD camera noise, as well as the design and fabrication of the telescope optical system. The focusing accuracy of astronomical telescopes is another important factor that affects the quality of the astronomical observations. The automatic focus system finds the best focus by moving the CCD imaging system and measures the image focus level by using a focus measure operator. The best focus occurs when the system reaches the position of maximum focus level. The imaging system moves according to the procedure of the used search algorithm.\par
Although adaptive optics (AO) has become a key to modern observations, the concept of the best focus still has a bit of importance, especially for telescopes that can't be provided by modern technology like adaptive optics. Because of its undeformable mirror and also due to the Adaptive optics' high cost. Adaptive optics is a system in which the received image is analyzed for distortion caused by seeing. \par
Several search algorithms are used for auto-focus systems, such as the famous Fibonacci search \cite{S. Rao.2009} and the modified Hill-Climbing search \cite{J. He.2003}. Besides, other well-known search algorithms like the Fibonacci search \cite{C. Batten.2000}, Binary search \cite{Y. Yao.2006}, and the Global search \cite{S. Rao.2009}. \par 
In this research work, the performance of several search algorithms is studied on astronomical images. The optimal search strategy will be used to develop an automatic focus system for Kottamia Astronomical Observatory (KAO).\par
The remainder of this paper is organized as follows: Section \ref{sec2} presents the popular search algorithms. In section \ref{sec3}, a description of the used focus measure operator is introduced. Section \ref{sec7} describes the used methodology. Finally, experimental results and conclusion are presented in sections \ref{sec4} and \ref{sec5}, respectively. \par

\section{Search Algorithms}\label{sec2}
The search algorithm is necessary for an imaging system that has a large number of focus positions. In the auto-focusing algorithm, the images focus levels are considered as the objective function of the search problem. The optimum solution of the search algorithm is supposed as the best focus position. In this section, the details of several search algorithms are presented. These algorithms were proposed with the assumption that the objective function having a unique maximum (unimodal). These algorithms can be applied to the maximization or minimization problem. But, in this research, the maximization one is investigated. \par

 As the system magnification increases, more severe blur is introduced, and the noise level increases accordingly. Furthermore, some sharpness measures cannot keep the required unimodal shape in such conditions. However, given the unimodal shape of the criterion function, the search algorithm can converge to the best focus location at a fast speed provided heuristic choices of the step magnitude \cite{Y. Yao.2006}. \par

\subsection{Global Search}
\begin{figure}[H]
  \centering
  \includegraphics[width=0.65\textwidth, height=9cm]{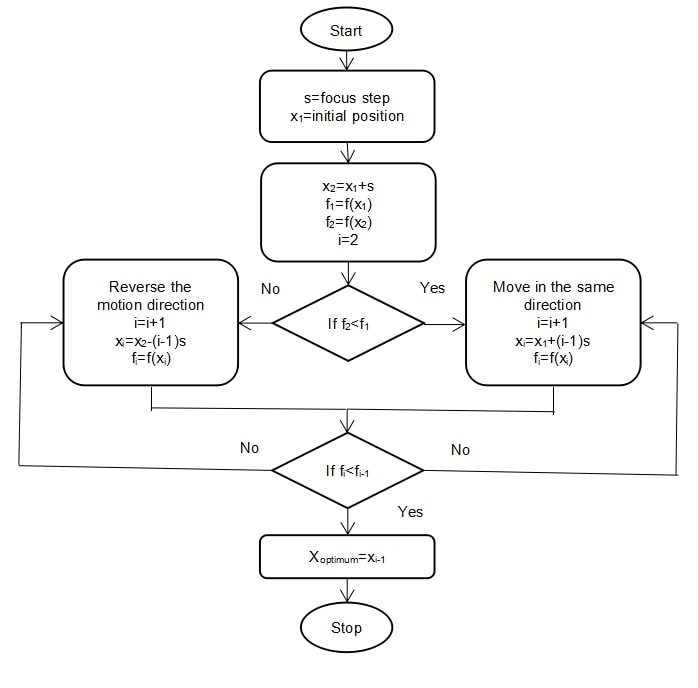}
  \caption{The Global search flow chart.}\label{fig217}
\end{figure}

The Global search \cite{S. Rao.2009} is the most straightforward method. It operates by passing through all focus positions using fixed step size until reaches the optimal (best) focus position. The step size used must be proportional to the desired accuracy. The search starts by acquiring an image from an initial focus position. Then, the focus level of the acquired image is calculated by applying a focus measure operator into the image. Afterward, the imaging system moves to the next position. The next position is determined by using the step size. The focus level of the acquired image at this position is also calculated. This procedure continues until a position that shows a decrease in the focus level value is reached. Finally, the search terminates, and the previous position is taken as the optimum one, as seen in figure \ref{fig217}. Such that $f_i$ is the image focus level at position $x_i$. The main disadvantage of this search is that it can only be applied in a narrow focus range case. That is because a large number of acquired images will be needed in case of a wide range. These large numbers will result in a large consumed time.\par

\subsection{Fibonacci Search}
The Fibonacci search \cite{C. Batten.2000,S. Rao.2009} successively narrows the search interval by using a known number of iterations which guarantees that the best focus position is found within the final focus range. The Fibonacci search depends on Fibonacci numbers. The Fibonacci number ($F_n$) is the sum of its two previous numbers in the Fibonacci series except for the first and second numbers in the series are equal to one, as in equation \ref{eq2.72} and \ref{eq2.73}.
\begin{equation}\label{eq2.72}
{F_n} = {\rm{ }}{F_n}_{ - 1} + {\rm{ }}{F_n}_{ - 2}
\end{equation}
\begin{equation}\label{eq2.73}
{F_0} = {\rm{ }}{F_1} = {\rm{ }}1
\end{equation}

\begin{figure}[H]
  \centering
  \includegraphics[width=0.65\textwidth]{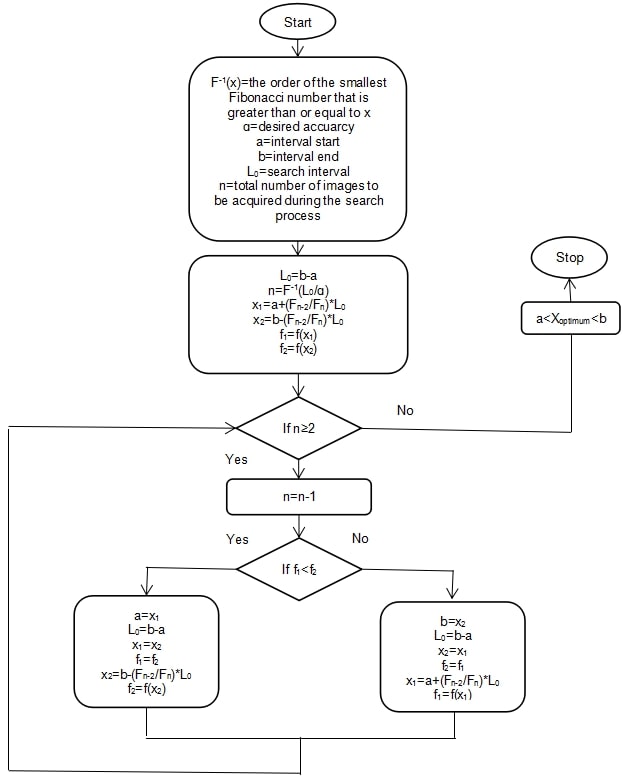}
  \caption{The Fibonacci search flow chart.}\label{fig218}
\end{figure}

By assuming that $L_0$ is the search interval ($L_0$=b-a), a is the interval start point, b is the interval endpoint, and n is the total number of images to be acquired during the search process. The search is executed as shown in figure \ref{fig218}.

\subsection{Binary Search}
\begin{figure}[H]
  \centering
  \includegraphics[width=12cm,height=8cm]{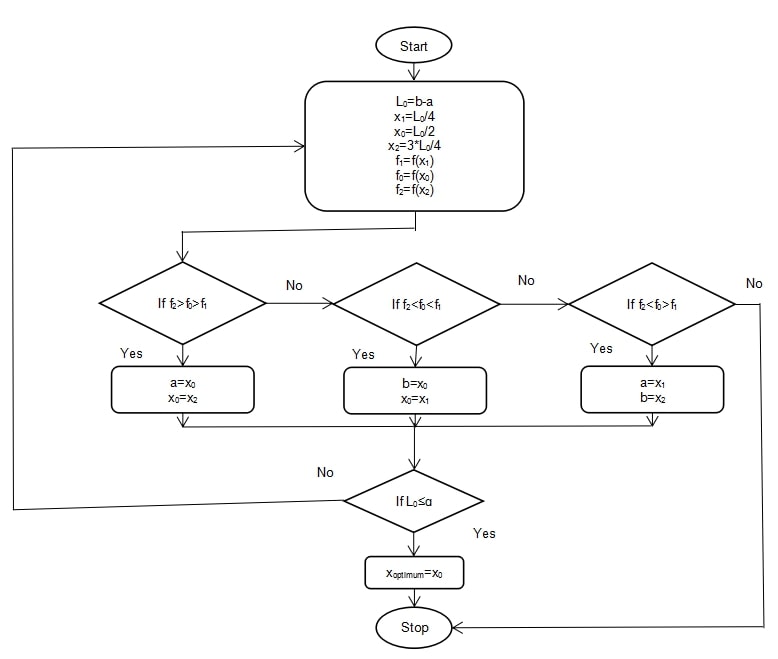}
  \caption{The Binary search flow chart.}\label{fig221}
\end{figure}

The Binary search \cite{S. Rao.2009,Y. Yao.2006}, which is named Interval Having method as well, is also an iterative search. It iteratively discards a part of the search interval, until it reaches the desired interval as in the Fibonacci search. But, in this method, half of the interval is discarded at each iteration. The Binary search theoretically needs three images to be acquired per iteration. The focus level of these three images determines which half is discarded. The Binary Search can be proceeded as in figure \ref{fig221}.

\subsection{Modified Fast Climbing Search (MFCS)}
\begin{figure}[H]
  \centering
  \includegraphics[width=5cm,height=7cm]{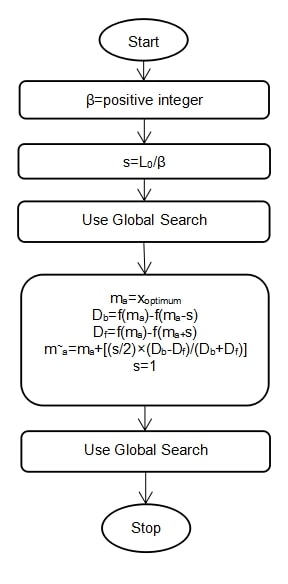}
  \caption{The MFCS flow chart.}\label{fig225}
\end{figure}

The Modified Fast Climbing Search (MFCS) \cite{J. He.2003} is an iteratively Global search using a different step value for each search region. In this method, the search interval is classified into two regions: out-of-focus and in-focus regions. In the out-of-focus region, the search proceeds with a large step size. While in the focused region (in-focus region), the search starts from initial peak position ($\widetilde {\mathop m\nolimits_a } $) using small focus step (s=1). The search uses a small step to refine the search result (i.e., more accurate result).
The Modified Fast Climbing search has been proposed as shown in figure \ref{fig225}.

\subsection{Subbarao Search}
\begin{figure}[H]
  \centering
  \includegraphics[width=3cm,height=9cm]{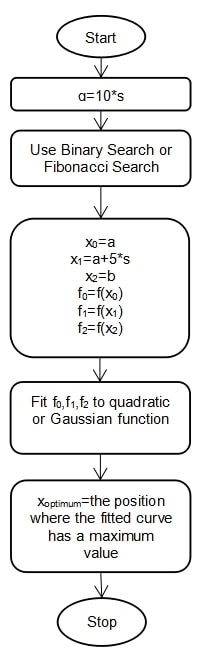}
  \caption{The Subbarao search flow chart.}\label{fig222}
\end{figure}

The Subbarao search \cite{M. Subbarao.1998} uses the Binary or Fibonacci search to reduce the search interval. Then, the final interval is equally separated, and three focus levels are fitted into a quadratic or Gaussian function. The position corresponding to the maximum of the fitted curve is considered as the best focus position, as shown in figure \ref{fig222}.

\section{Focus Measure Operator}\label{sec3}
The image focus level is measured by applying a focus measure operator into the image.  In \cite{I. Helmy.2020}, a comparative study for different nineteen operators applied to five sequences of astronomical star-clusters observations is introduced. The results clarify that the Normalized Variance focus measure has the best overall performance. Thus, in this paper, we investigate the same operator with the different search algorithms. \par

By dividing the image variance with the image mean ($\bar{I}$), the Normalized Variance is advised to compensate for the differences in the average image brightness among different images. The Normalized Variance focus measure ($\rm{FM}$) is given in equation \ref{eq2.25}, where I(x,y) is the pixel intensity at a coordinate (x,y), and $\rm{M}\times \rm{N}$ pixels is the image size.\par
\begin{equation}\label{eq2.25}
{\rm{FM}} = \frac{{\frac{1}{{{\rm{MN}}}}\mathop \sum \limits_{\rm{x=1}}^{\rm{M}} \mathop \sum \limits_{\rm{y=1}}^{\rm{N}}{{\left( {{\rm{I}}\left(
{{\rm{x}},{\rm{y}}} \right) - {\rm{\bar{I}}}} \right)}^2}}}{{\rm{\bar{I}}}}
\end{equation}

\section{Methodology}\label{sec7}
The search algorithms mentioned in section \ref{sec2} are applied to five sequences of star-clusters observations. The sequences are observed by an imaging system based on the 74-inches telescope of the Kottamia astronomical observatory and CCD camera system \cite{Y. Azzam.2008}. The observed star-clusters are M103, N6793, N7067, N7788, and N7789. The sequences are observed during good seeing conditions. It contains in-focus and out-of-focus frames. Table \ref{tab5.9} shows the details of the astronomical observations. \par
The seeing conditions prevailing at the site is around 2 arcsec on average \cite{Y. Azzam.2008}. The Newtonian Plate is 22.53 arcsec/mm, and the Imaging area is 27.6 mm by 27.6 mm. As a consequence, the pixel is 0.30 arcsec ( (22.53$\times$27.6)/2048 ). Thus. the average 2 arcsec is around 6.5 pixels. Yet, the observation was acquired due to good seeing condition, and the Full-Width Half Maximum (FWHM) for the best-focus image doesn't exceed 5 pixels, as you can see in table \ref{tab5.9}. \par 
Unfortunately, we don't have a Differential Image Motion Monitor (DIMM). Besides, the Polaris telescope didn't work properly during the observation interval. Finally, the values of the average FWHM of each sequence has inserted in table \ref{tab5.9}. \par

\begin{table*}
  \caption{The summary of the used star-clusters.}\label{tab5.9}
  \centering
\resizebox{\textwidth}{0.08\textwidth}{
\begin{tabular}{@{}cccccccc@{}} \toprule
	Star-Cluster Name & Start Position ($\mu$m)  & End Position ($\mu$m)  & Range ($\mu$m)  & Step Size ($\mu$m) & Number of Frames &  Exposure Time / Frame (seconds) & Average FWHM (Pixels) \\ \colrule
        M103 & 48300 & 54800 & 6600 & 100 & 66 & 60  & 4.15  \\ 
        N6793 & 49000 & 55000 & 6100 & 100 & 61 & 60  & 4.279 \\   
        N7067 & 49000 & 54400 & 5500 & 100 & 55& 60  & 3.68  \\ 
        N7788 & 48600 & 54600 & 6100 & 100 & 61 & 60 & 4.78 \\ 
        N7789 & 48000 & 55000 & 7100 & 100 & 71& 60 &  3.89  \\
  \botrule
  \end{tabular}}
\end{table*}

The impact of the atmospheric conditions (seeing) can be measured by the Full-Width Half Maximum (FWHM).  In this work, we depend on observations acquired using the CCD camera. The evaluation of either the search algorithms or the focus measure operator relies on the image focus level, which is estimated using the FWHM. The FWHM is computed using a well-known program, called MAXIM DL. \par

\begin{figure}
\centering
\begin{subfigure}{.5\textwidth}
  \centering
  \includegraphics[width=8cm,height=4.75cm]{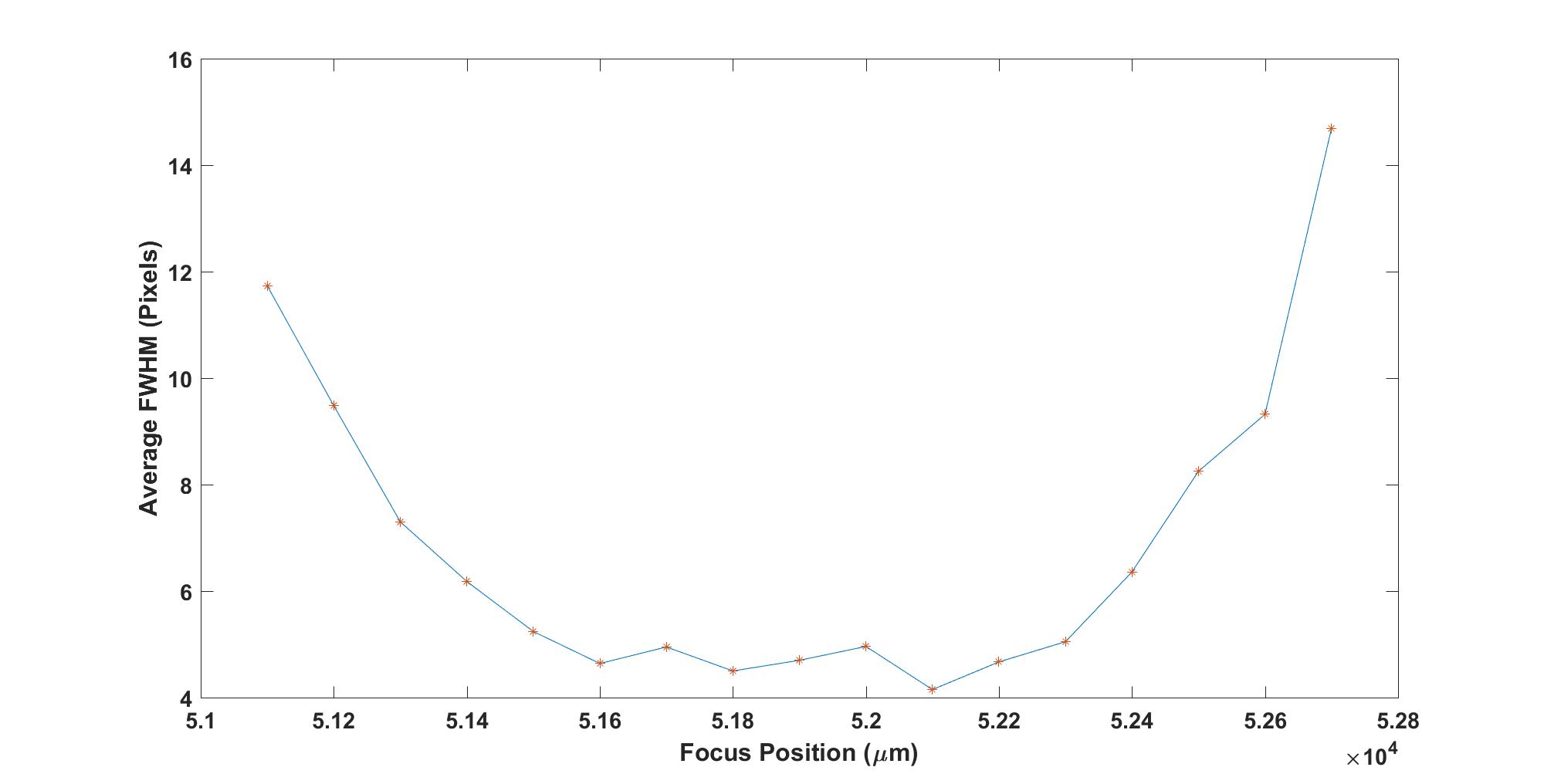}
  \caption{The FWHM distribution of the sequence M103.}
  \label{fig20a}
\end{subfigure}%
\hspace*{8pt}
\begin{subfigure}{.5\textwidth}
  \centering
  \includegraphics[width=8cm,height=4.75cm]{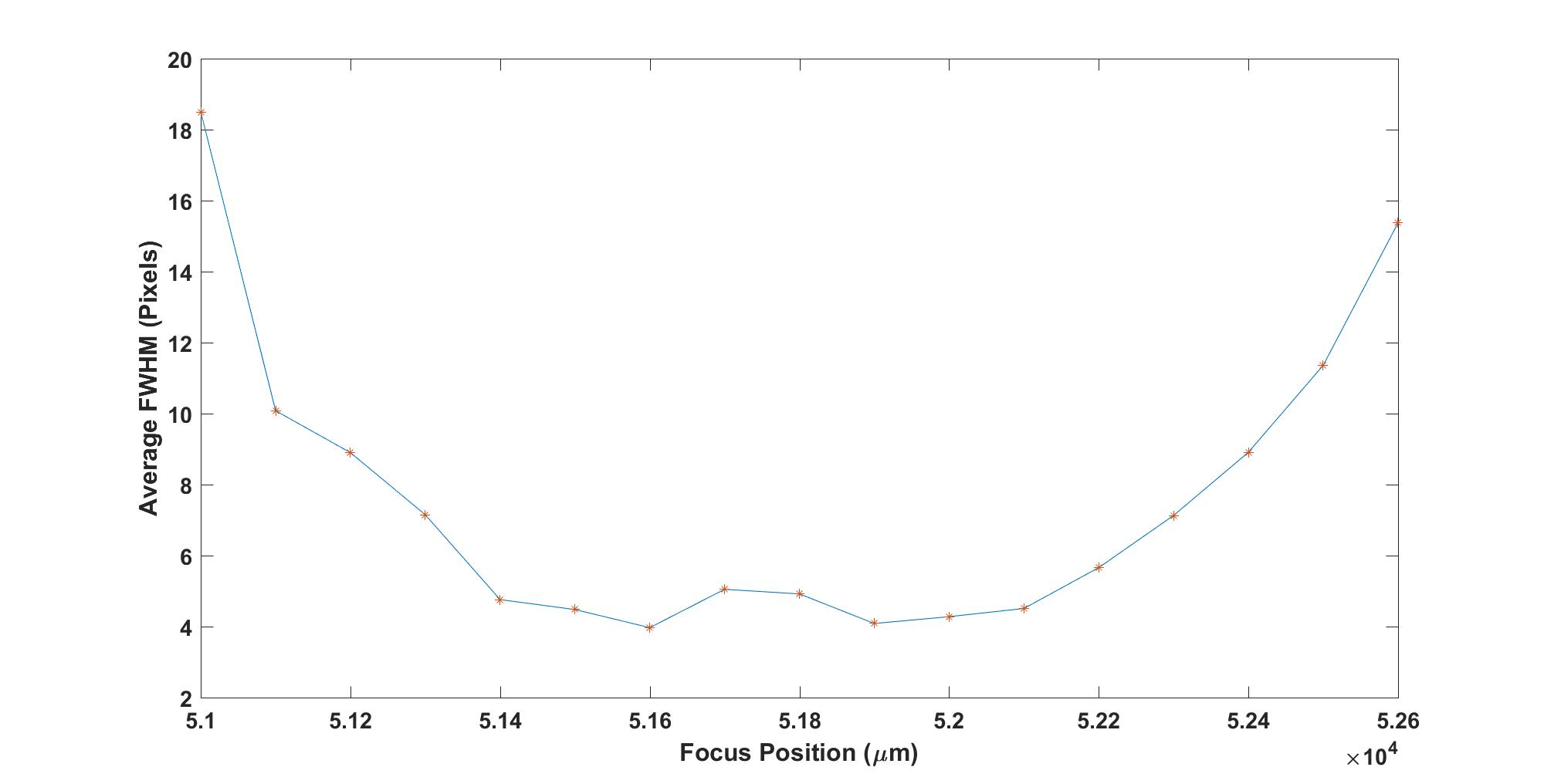}
  \caption{The FWHM distribution of the sequence N6793.}
  \label{fig20b}
\end{subfigure}

\begin{subfigure}{.5\textwidth}
  \centering
  \includegraphics[width=8cm,height=4.75cm]{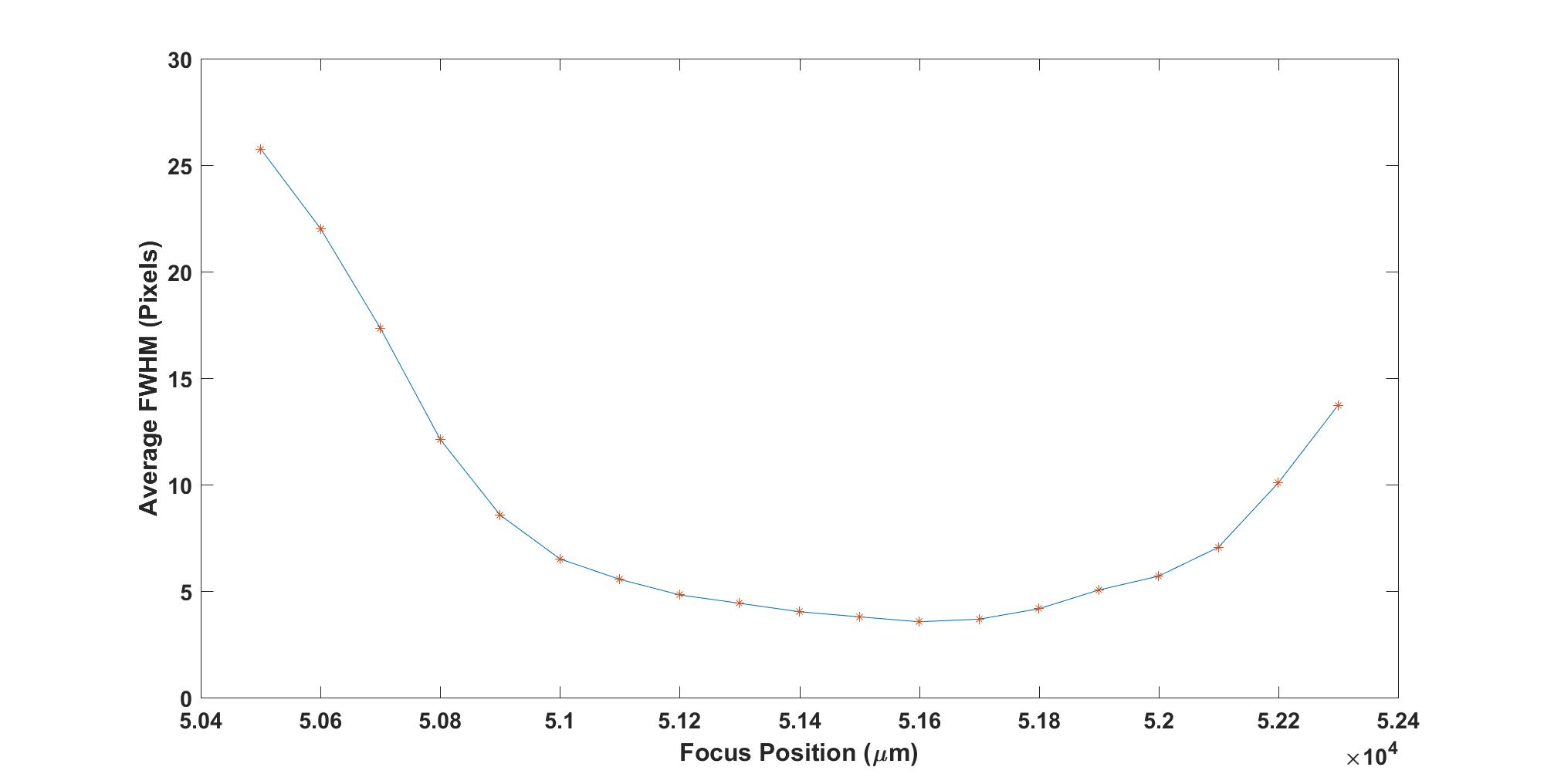}
  \caption{The FWHM distribution of the sequence N7067.}
  \label{fig20c}
\end{subfigure}%
\hspace*{8pt}
\begin{subfigure}{.5\textwidth}
  \centering
  \includegraphics[width=8cm,height=4.75cm]{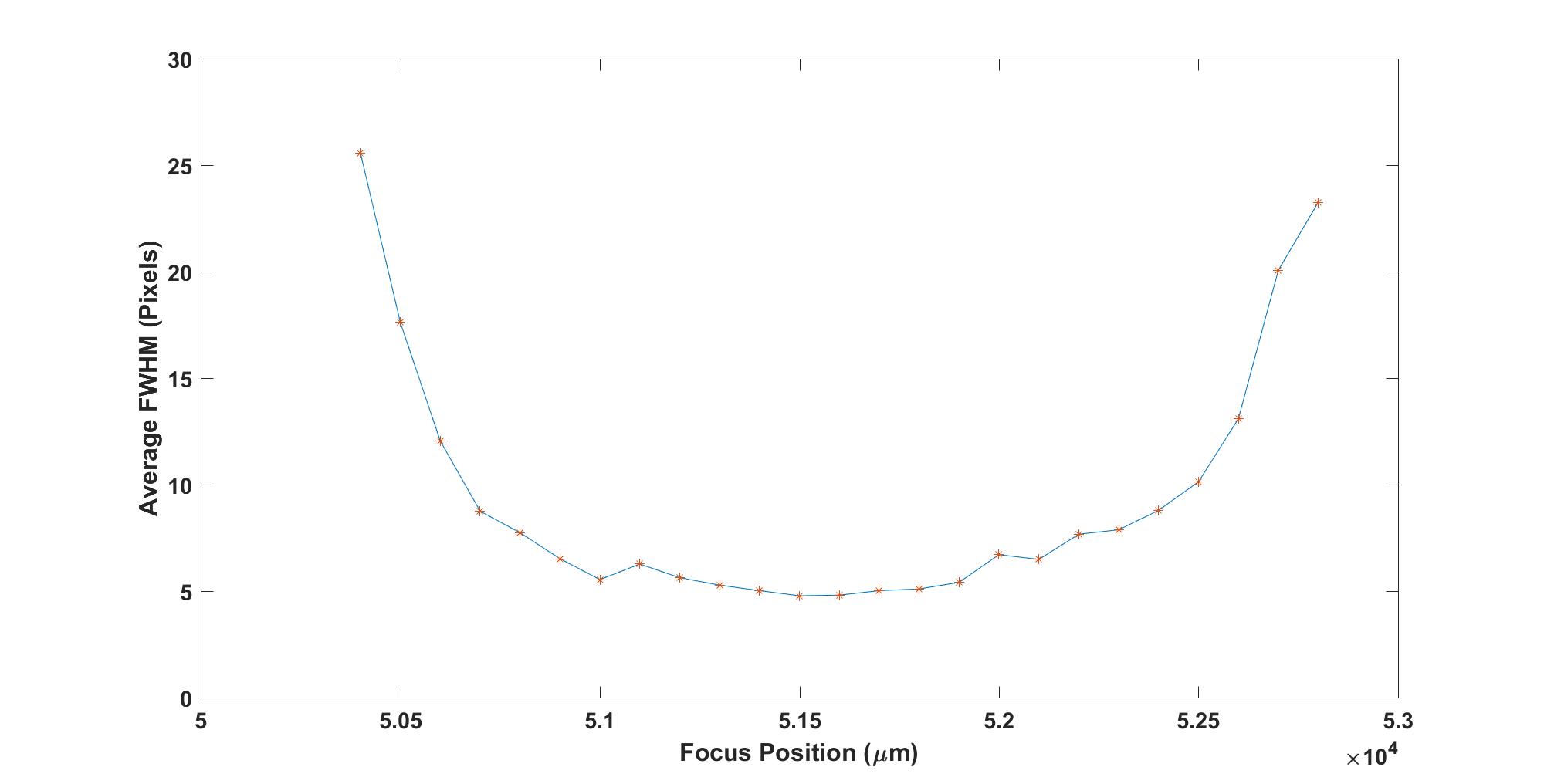}
  \caption{The FWHM distribution of the sequence N7788.}
  \label{fig20d}
\end{subfigure}

\begin{subfigure}{.5\textwidth}
  \centering
  \includegraphics[width=8cm,height=4.75cm]{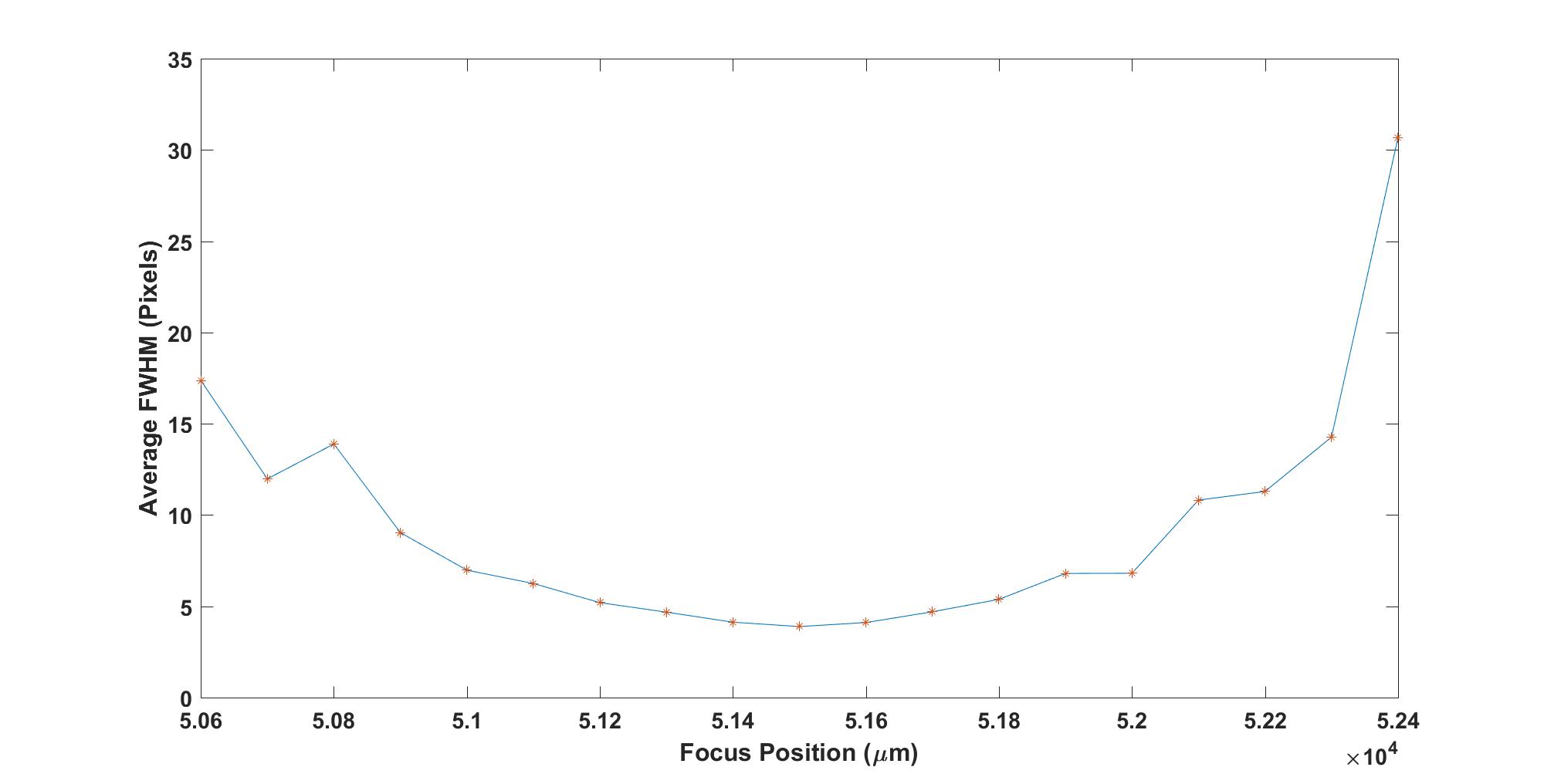}
  \caption{The FWHM distribution of the sequence N7789.}
  \label{fig20e}
\end{subfigure}

	\caption{The FWHM distributions of the different sequences. (a) : The FWHM distribution of the sequence M103. (b) : The FWHM distribution of the sequence N6793. (c) : The FWHM distribution of the sequence N7067. d) : The FWHM distribution of the sequence N7788. e) : The FWHM distribution of the sequence N7789. }\label{fig2}
\end{figure}

Figure \ref{fig20a}, \ref{fig20b}, \ref{fig20c}, \ref{fig20d}, and \ref{fig20e} shows the distribution of the average FWHM value of the sequence M103, N6793, N7067, N7788, and N7789, respectively. The FWHM is calculated for an isolated unsaturated star. The FWHM is the best focus measure, however, its main disadvantage that can't be applied outside the focus region, which makes the search processing is limited, as well as requiring previous experience during the observation process. \par

\begin{figure}[H]
\centering
\begin{subfigure}{.5\textwidth}
  \centering
  \includegraphics[width=4cm,height=4cm]{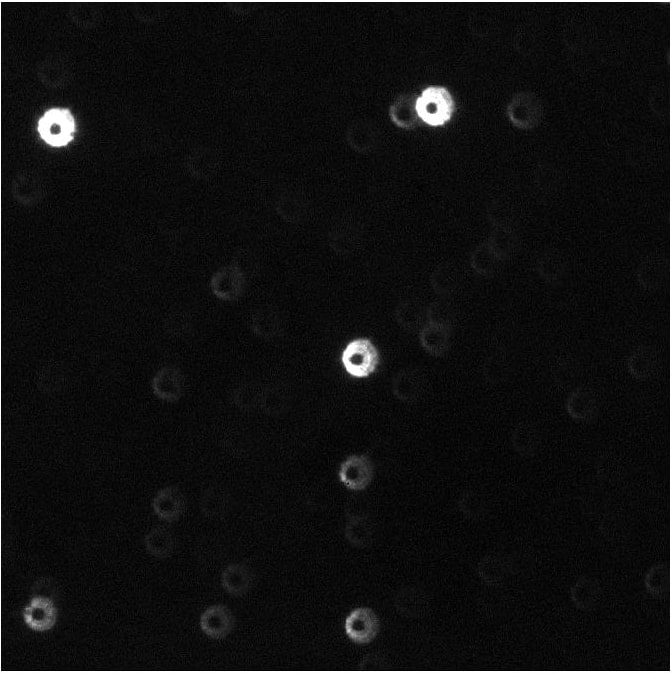}
  \caption{Out of Focus.}
  \label{fig31a}
\end{subfigure}%
\hspace*{8pt}
\begin{subfigure}{.5\textwidth}
  \centering
  \includegraphics[width=4cm,height=4cm]{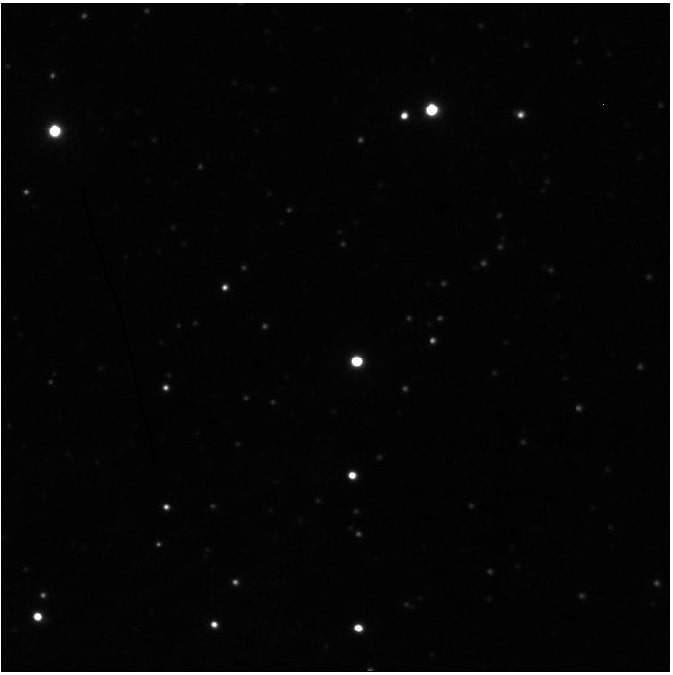}
  \caption{Best Focus.}
  \label{fig31b}
\end{subfigure}
\vspace*{8pt}
	\caption{A representative example of sequence M39. (a) : Out of Focus. (b) : Best Focus.}\label{fig31}
\end{figure}

The images are of size 2048 $\times$ 2048 pixels. A representative example of images at different positions is shown in figure \ref{fig31a} and \ref{fig31b}. The performance evaluation of the various search algorithms is based on two criteria, which are: 1) the number of steps the search process needs until reaches the optimum solution. 2) the accuracy error of the optimum solution from the best solution. The search algorithms are ranked according to their scores. The scores are calculated, as expressed in equation \ref{eq.31}. This equation has been found to achieve the compromise between the number of steps and error.\par

\begin{equation}\label{eq.31}
{\Large Score_{i}= \frac{max(s_{i})\times min(a_{i})}{s_{i} \times a_{i}}}
\end{equation}

Where:
i=1,...,n, $Score_i$ is the score of the search algorithm, n is the number of the search algorithms, $a_i$ is the accuracy error of the search algorithm, $s_i$ is the number of steps of the algorithm.\par

\begin{figure}[H]
\centering
\begin{subfigure}{.5\textwidth}
  \centering
  \includegraphics[width=8cm,height=4.75cm]{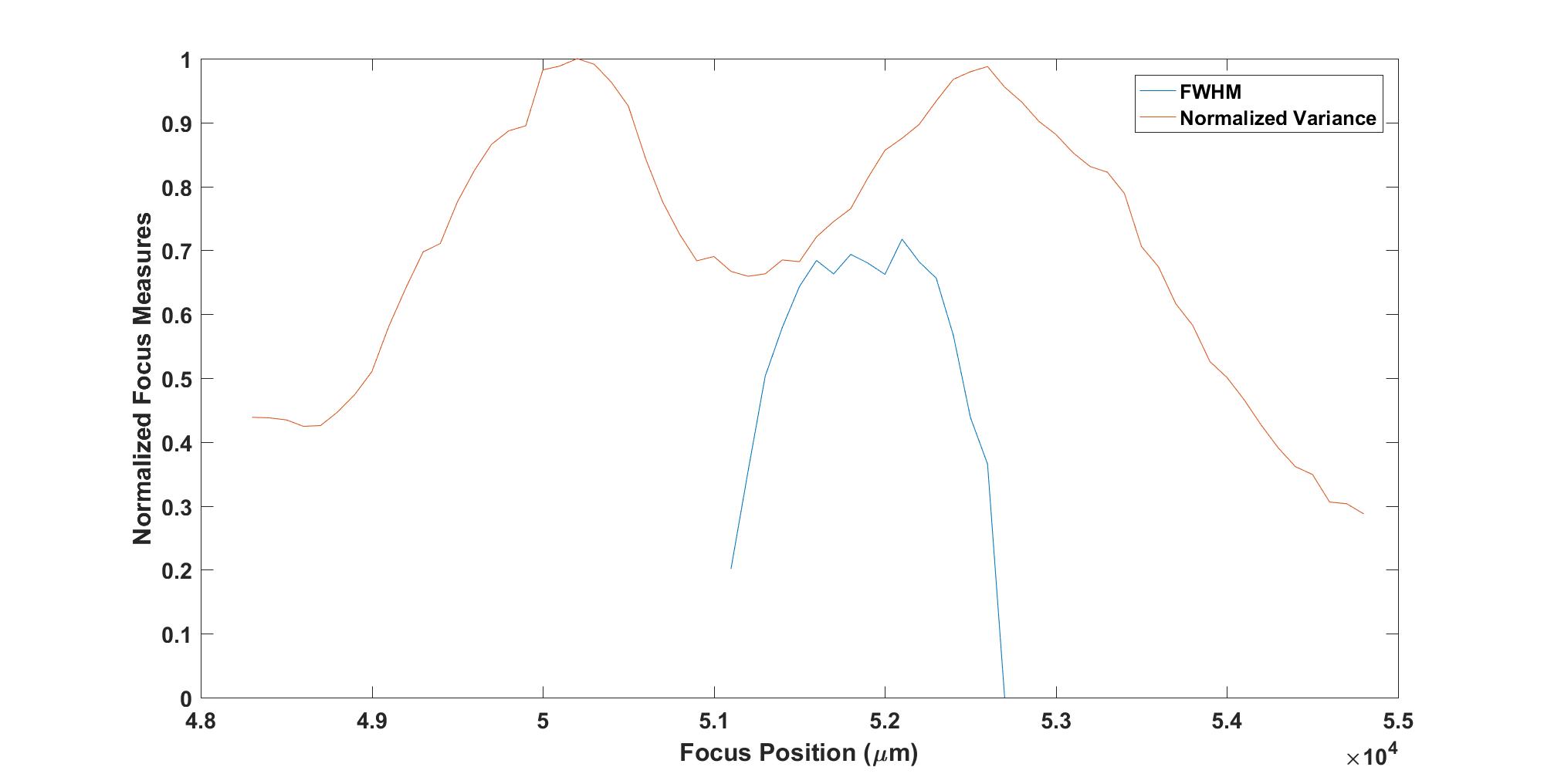}
  \caption{The distribution of the Normalized Variance focus measure with the FWHM of the sequence M103.}
  \label{fig30a}
\end{subfigure}%
\hspace*{8pt}
\begin{subfigure}{.5\textwidth}
  \centering
  \includegraphics[width=8cm,height=4.75cm]{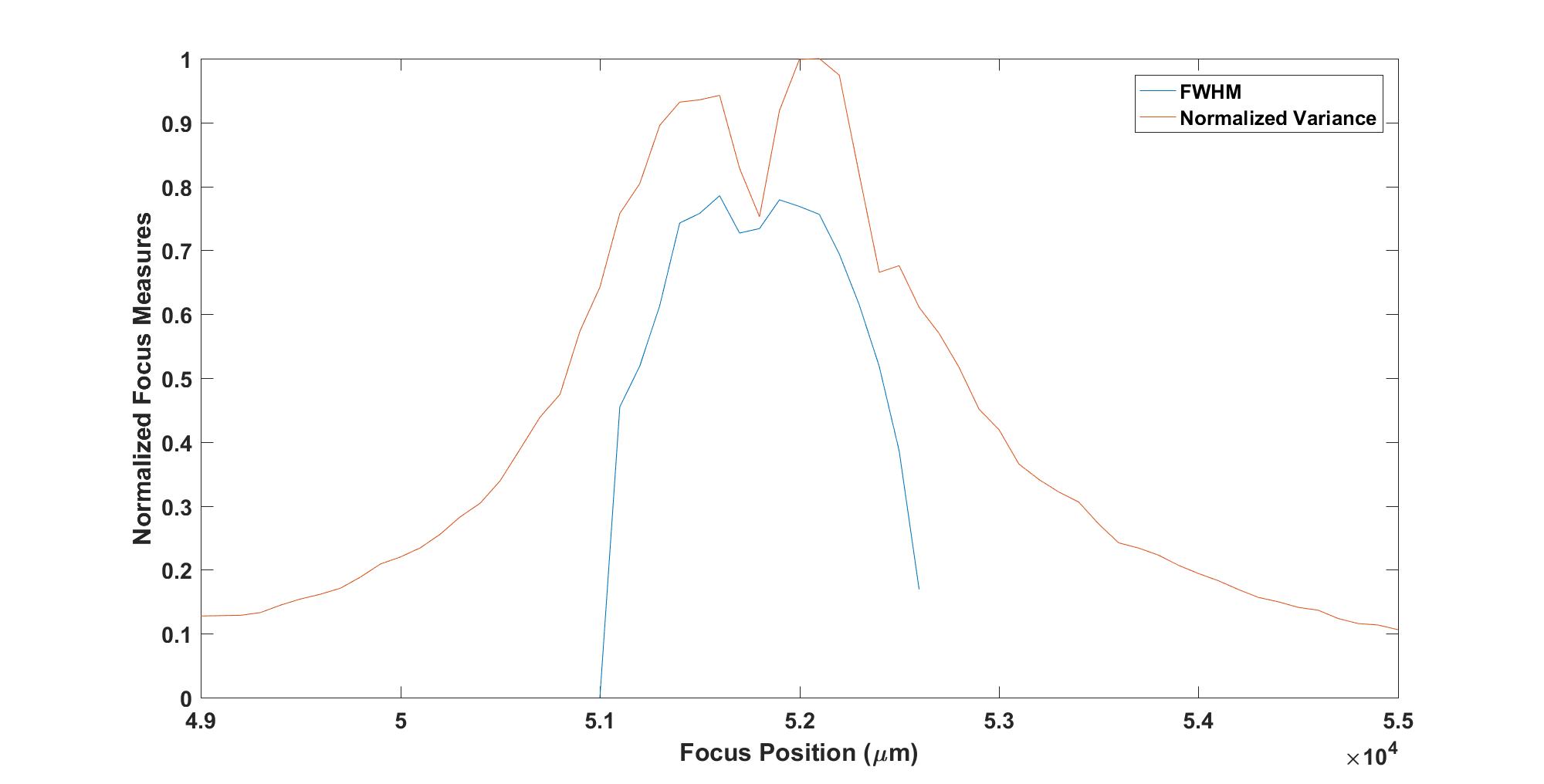}
  \caption{The FWHM distribution The distribution of the Normalized Variance focus measure with the FWHM of the sequence N6793.}
  \label{fig30b}
\end{subfigure}

\begin{subfigure}{.5\textwidth}
  \centering
  \includegraphics[width=8cm,height=4.75cm]{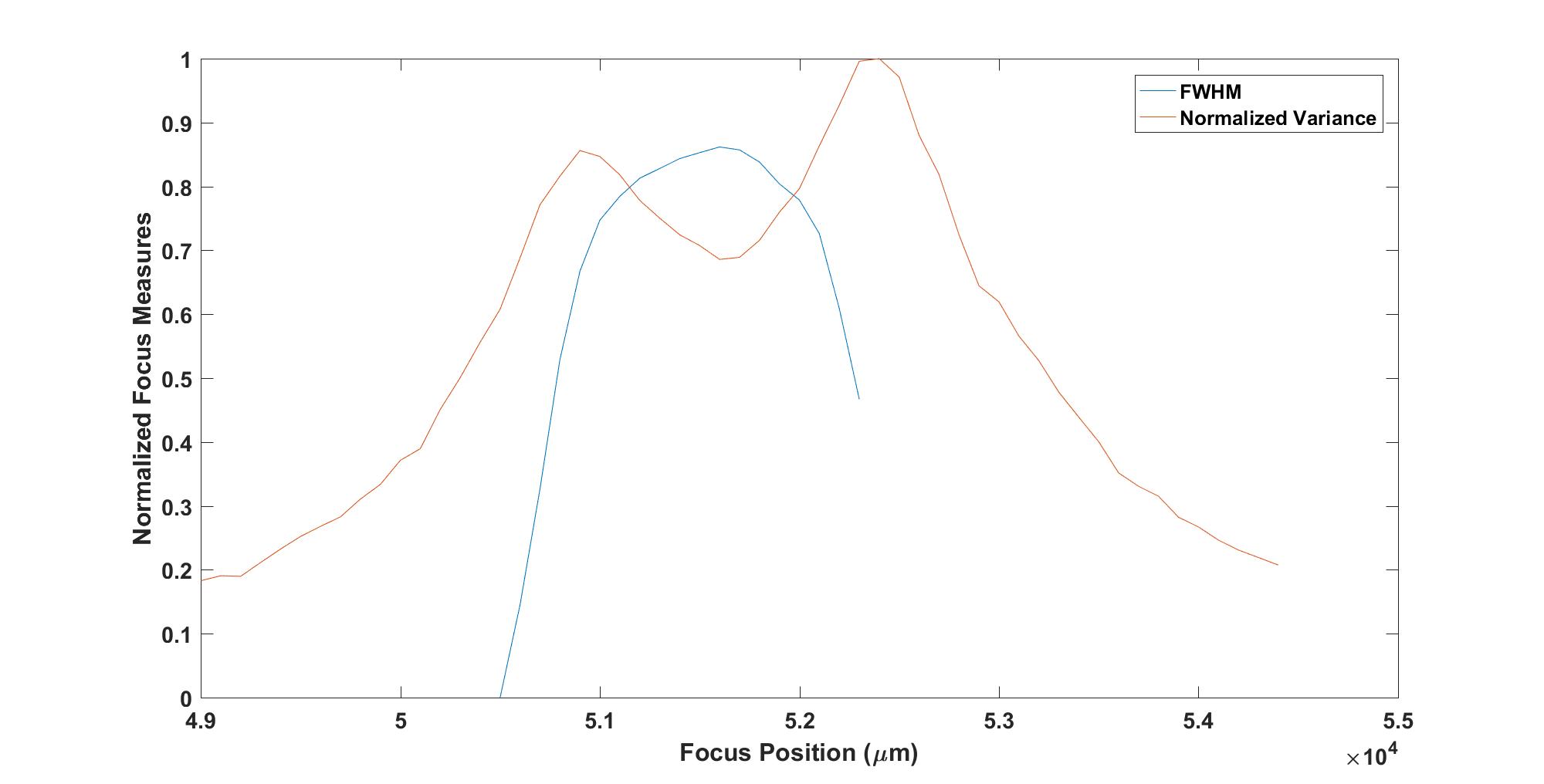}
  \caption{The distribution of the Normalized Variance focus measure with the FWHM of the sequence N7067.}
  \label{fig30c}
\end{subfigure}%
\hspace*{8pt}
\begin{subfigure}{.5\textwidth}
  \centering
  \includegraphics[width=8cm,height=4.75cm]{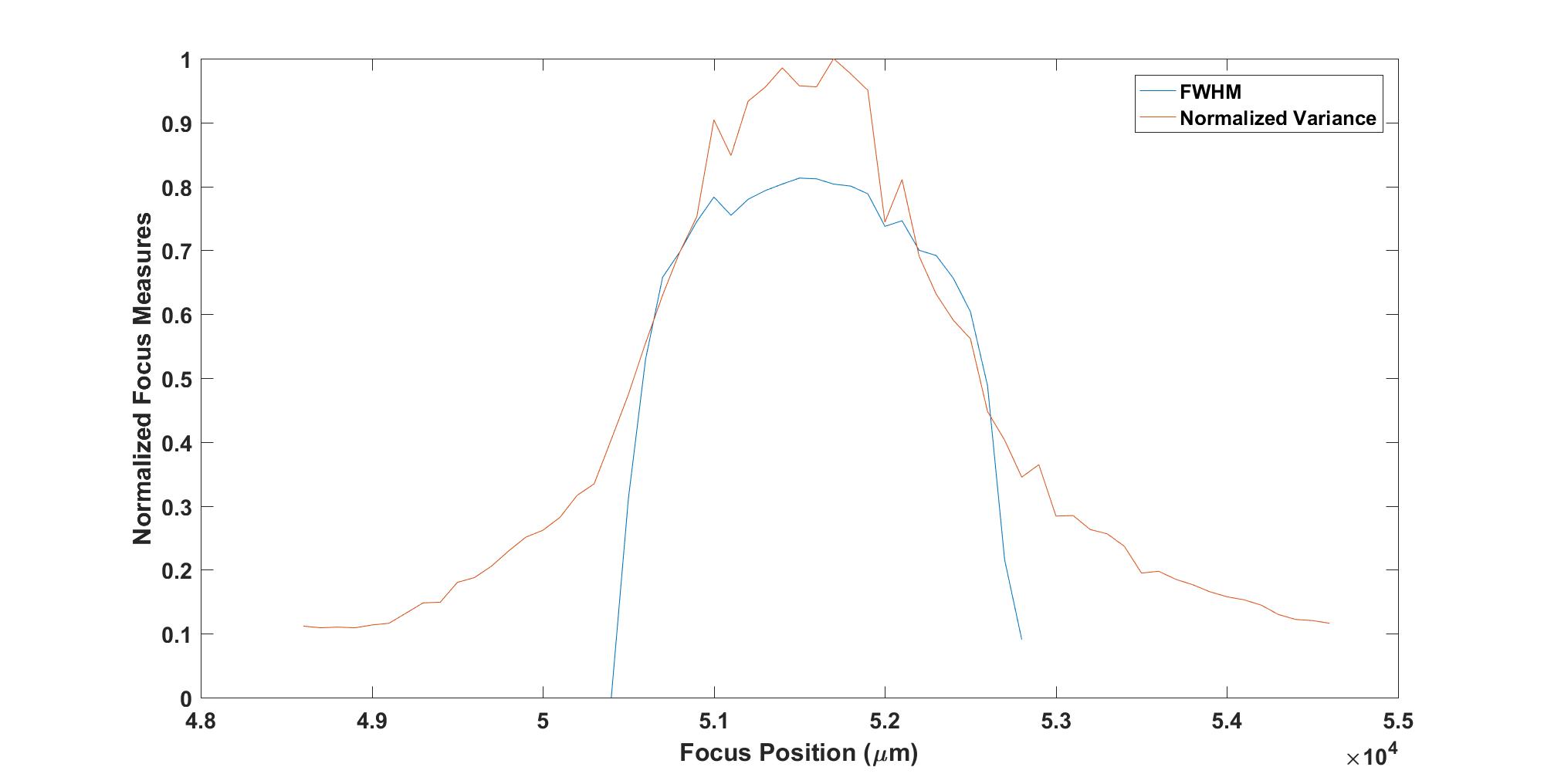}
  \caption{The distribution of the Normalized Variance focus measure with the FWHM of the sequence N7788.}
  \label{fig30d}
\end{subfigure}

\begin{subfigure}{.5\textwidth}
  \centering
  \includegraphics[width=8cm,height=4.75cm]{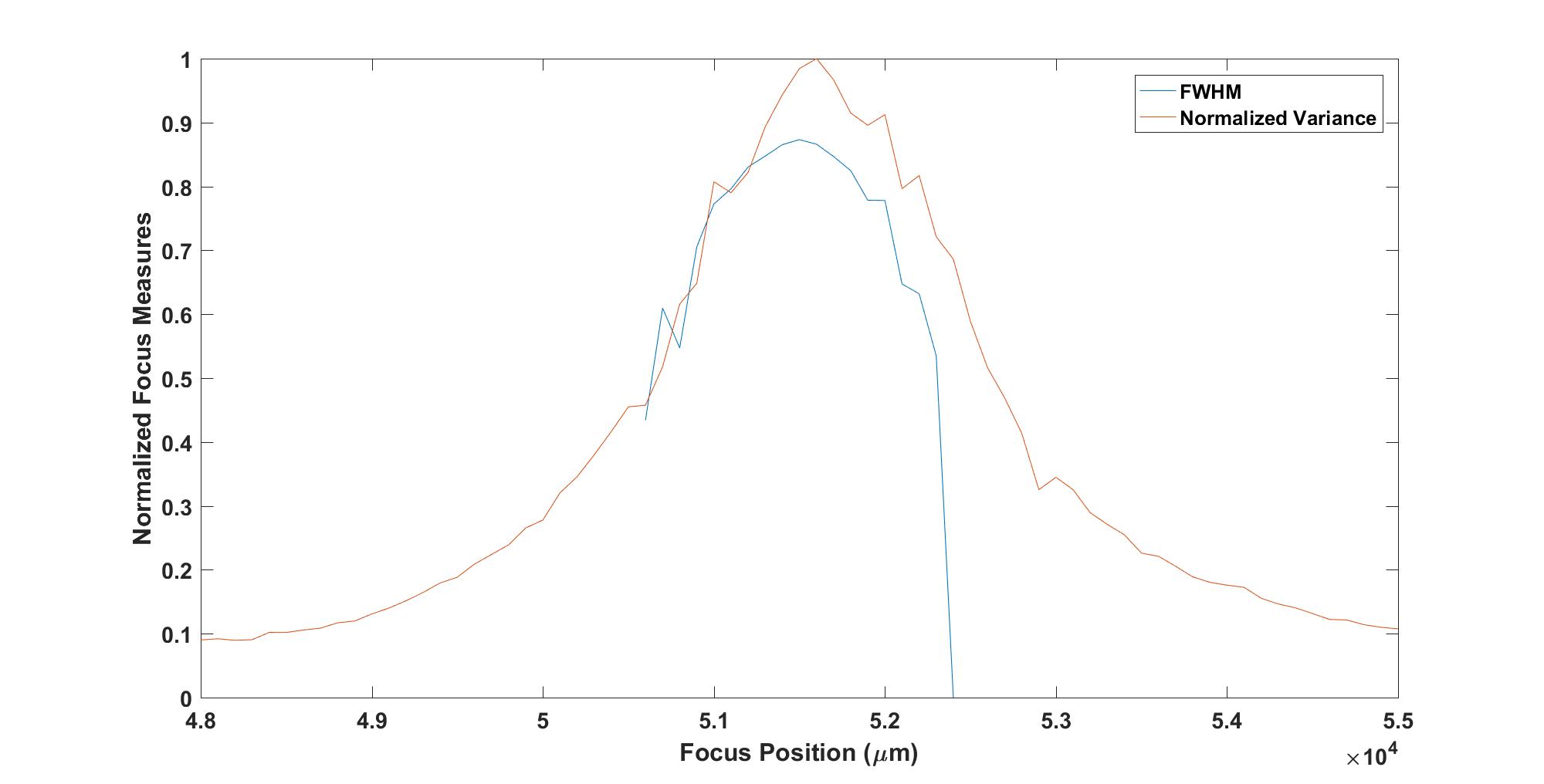}
  \caption{The distribution of the Normalized Variance focus measure with the FWHM of the sequence N7789.}
  \label{fig30e}
\end{subfigure}

	\caption{The distributions of the Normalized Variance focus measure with the FWHM of the different sequences. (a) : The distribution of the Normalized Variance focus measure with the FWHM of the sequence M103. (b) : The FWHM distribution of the sequence N6793. (c) : The distribution of the Normalized Variance focus measure with the FWHM of the sequence N7067. d) : The distribution of the Normalized Variance focus measure with the FWHM of the sequence N7788. e) : The distribution of the Normalized Variance focus measure with the FWHM of the sequence N7789. }\label{fig2}
\end{figure}

Figure \ref{fig30a}, \ref{fig30b}, \ref{fig30c}, \ref{fig30d}, and \ref{fig30e} presents the distribution of the Normalized Variance focus measure with the FWHM of the sequence M103, N6793, N7067, N7788, and N7789, respectively. The FWHM has the privilege of a focus operator. However, The used focus measure (Normalized Variance) is selected as an optimum focus measure according to the study in \cite{I. Helmy.2020}. As a consequence, we used it as the guide during the search process, which is the prime focus of this research. \par

\section{Experimental Results and Discussions}\label{sec4}

Table \ref{tab5.10} shows the results of the algorithms applied to sequence M103. The results show that the Binary is the best algorithm. Whereas, Global is the worst operator. That is because it is suitable only for the small search interval and ideal conditions. Otherwise, it will require a large number of steps or fall in the first local peak.

\begin{table}[H]
  \caption{The results of the search algorithms comparison applied to M103.}\label{tab5.10}
  \centering
\begin{tabular}{@{}cccc@{}} \toprule
	Search Name & Number of steps & Accuarcy Error ($\mu$m) & Score \\ \colrule
        Binary & 17 & 162 & 1.5294 \\ 
        Modified Fast Climbing & 12 & 302 & 1.1623 \\ 
        Fibonacci & 10 & 363 & 1.1603 \\ 
        Subbarao-Binary & 26 & 167 & 0.9701 \\ 
        Subbarao-Fibonacci & 17 & 351 & 0.7059 \\ 
        Global & 2 & 3800 & 0.5542 \\ 
  \botrule
  \end{tabular}

\end{table}

In table \ref{tab5.11}, the results of algorithms applied to sequence N6793 are presented. The results show the Binary is the best algorithm. As same as the M103 case, Global is the worst operator, and the Subbarao-Fibonacci and Fibonacci have almost the same rank.

\begin{table}[H]
  \caption{The results of the search algorithms comparison applied to N6793.}\label{tab5.11}
  \centering
\begin{tabular}{@{}cccc@{}} \toprule
	Search Name & Number of steps & Accuarcy Error ($\mu$m) & Score \\ \colrule
        Binary & 18 & 94 & 1.4894 \\ 
        Subbarao-Binary & 24 & 90 & 1.1667 \\ 
        Modified Fast Climbing & 13 & 207 & 0.9365 \\ 
        Fibonacci & 10 & 437 & 0.5767 \\ 
        Subbarao-Fibonacci & 17 & 438 & 0.3384 \\ 
        Global & 28 & 400 & 0.225 \\ 
  \botrule
  \end{tabular}

\end{table}

In table \ref{tab5.12}, the results of search algorithms applied to sequence N7067 are displayed. The results show the Binary is the best algorithm. However, the Modified Fast Climbing is the worst operator, and the Subbarao-Fibonacci and Fibonacci have almost as the M103 and N6793 cases.

\begin{table}[H]
  \caption{The results of the search algorithms comparison applied to N7067.}\label{tab5.12}
  \centering
\begin{tabular}{@{}cccc@{}} \toprule
	Search Name & Number of steps & Accuarcy Error ($\mu$m) & Score \\ \colrule
        Binary & 15 & 85 & 1.5059 \\ 
        Subbarao-Binary & 24 & 80 & 1 \\ 
        Fibonacci & 9 & 737 & 0.2895 \\ 
        Global & 3 & 2600 & 0.2462 \\ 
        Subbarao-Fibonacci & 17 & 553 & 0.2042 \\ 
        Modified Fast Climbing & 15 & 673 & 0.1902 \\ 
  \botrule
  \end{tabular}
\end{table}

Table \ref{tab5.13} presents the results of algorithms applied to sequence N7788. The results show the Fibonacci is the best algorithm. Whereas Global is the worst operator. It should be mentioned that the Binary score is almost the same as in the previous results (M103, N6793, and N7067).

\begin{table}[H]
  \caption{The results of the search algorithms comparison applied to N7788.}\label{tab5.13}
  \centering
\begin{tabular}{@{}cccc@{}} \toprule
	Search Name & Number of steps & Accuarcy Error ($\mu$m) & Score \\ \colrule
        Fibonacci & 10 & 135 & 1.7067 \\ 
        Binary & 15 & 101 & 1.5208 \\ 
        Subbarao-Binary & 24 & 96 & 1 \\ 
        Subbarao-Fibonacci & 17 & 143 & 0.9478 \\ 
        Modified Fast Climbing & 11 & 239 & 0.8764 \\ 
        Global & 2 & 2900 & 0.3972 \\ 
  \botrule
  \end{tabular}
\end{table}

Table \ref{tab5.14} displays the results of algorithms applied to sequence N7789. The results show the Fibonacci operator is the best measure. However, Global is the worst operator. The results of this sequence are not similar to the previous cases. Such that, the Binary has the superior to the other members, while in that case reserve the fourth rank. Yet, Global has almost the same rank as the previous cases.

\begin{table}[H]
  \caption{The results of the search algorithms comparison applied to N7789.}\label{tab5.14}
  \centering
\begin{tabular}{@{}cccc@{}} \toprule
	Search Name & Number of steps & Accuarcy Error ($\mu$m) & Score \\ \colrule
        Fibonacci & 10 & 39 & 2.6 \\ 
        Subbarao-Fibonacci & 18 & 44 & 1.2803 \\ 
        Modified Fast Climbing & 13 & 78 & 1 \\ 
        Binary & 17 & 110 & 0.5422 \\ 
        Subbarao-Binary & 26 & 105 & 0.3714 \\ 
        Global & 3 & 3400 & 0.0994 \\ 
  \botrule
  \end{tabular}
\end{table}

Table \ref{tab5.15} shows the overall score of all search algorithms for the five astronomical observations. The scores are computed by averaging the scores of all algorithms. The results clarify that the Binary has the best overall score of 1.3175. The second rank has been reserved by Fibonacci with a score of 1.2666. Finally, Subbarao-Binary and Modified Fast Climbing have almost the same scores of 0.9016 and 0.8331, respectively.

\begin{table}[H]
  \caption{The results summary for the different search algorithms.}\label{tab5.15}
  \centering
\begin{tabular}{@{}ccccccc@{}} \toprule
	Search Name & M103 & N6793 & N7067 & N7788 & N7789 & Overall Score \\ \colrule
        Binary & 1.5294 & 1.4894 & 1.5059 & 1.5208 & 0.5422 & 1.3175 \\ 
        Fibonacci & 1.1603 & 0.5767 & 0.2895 & 1.7067 & 2.6 & 1.2666 \\ 
        Subbarao-Binary & 0.9701 & 1.1667 & 1 & 1 & 0.3714 & 0.9016 \\ 
        Modified Fast Climbing & 1.1623 & 0.9365 & 0.1902 & 0.8764 & 1 & 0.8331 \\ 
        Subbarao-Fibonacci & 0.7059 & 0.3384 & 0.2042 & 0.9478 & 1.2803 & 0.6953 \\ 
        Global & 0.5542 & 0.225 & 0.2462 & 0.3972 & 0.0994 & 0.3044 \\ 
  \botrule
  \end{tabular}
\end{table}

Table \ref{tab5.15} shows that the best search algorithms for the case study (large search interval) are the Binary and the Fibonacci. However, time plays a key role in the real-time observation process. As a consequence, astronomers, in the science runs, often start searching depend on a previous estimated focus model that varies with the tube temperature or through the astronomer experience. \par

\begin{figure}[H]
\centering
\begin{subfigure}{.5\textwidth}
  \centering
  \includegraphics[width=8cm,height=5.5cm]{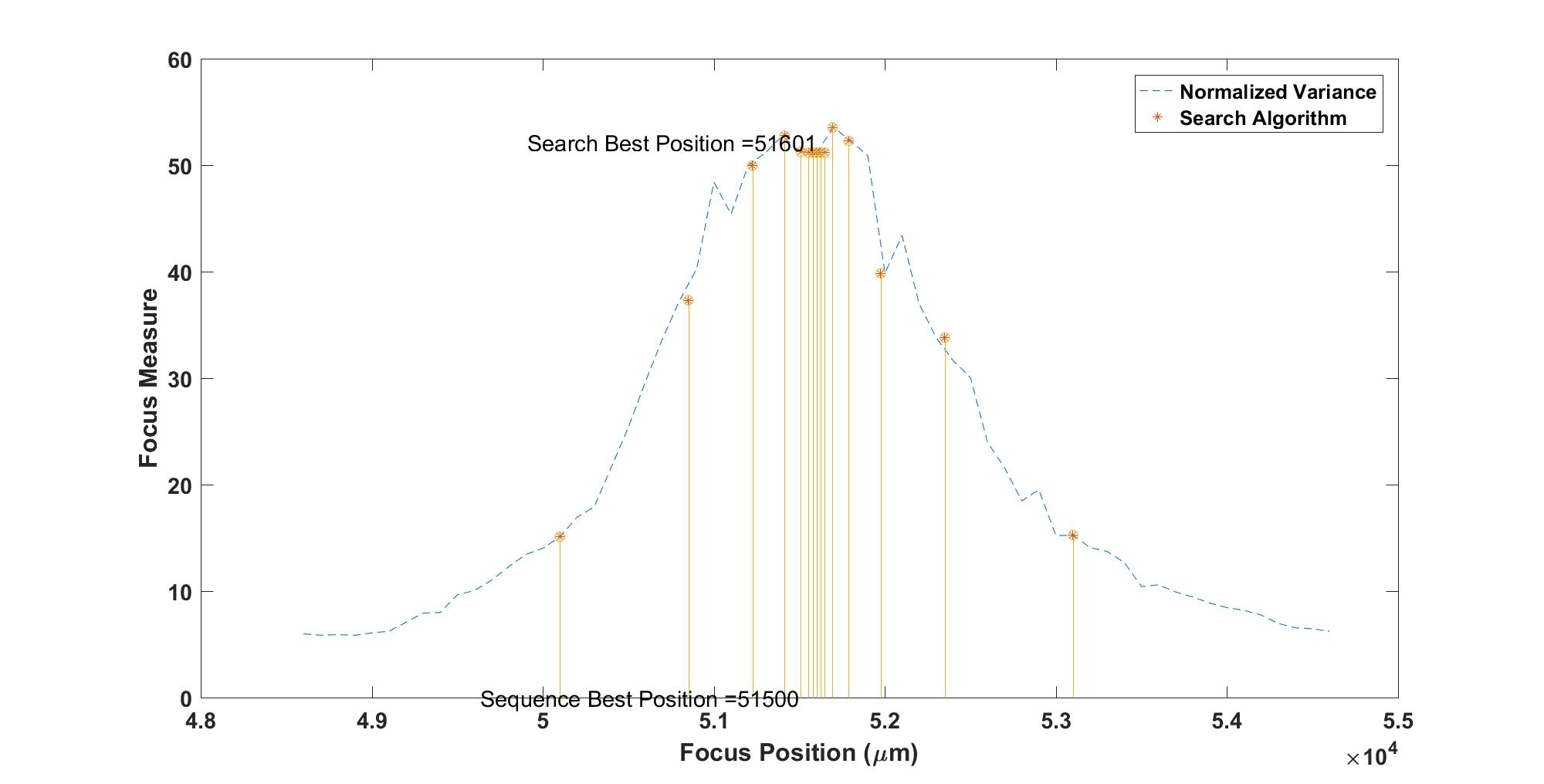}
  \caption{The steps of the Binary search algorithm of the sequence N7788.}
  \label{fig7a}
\end{subfigure}%
\hspace*{8pt}
\begin{subfigure}{.5\textwidth}
  \centering
  \includegraphics[width=8cm,height=5.5cm]{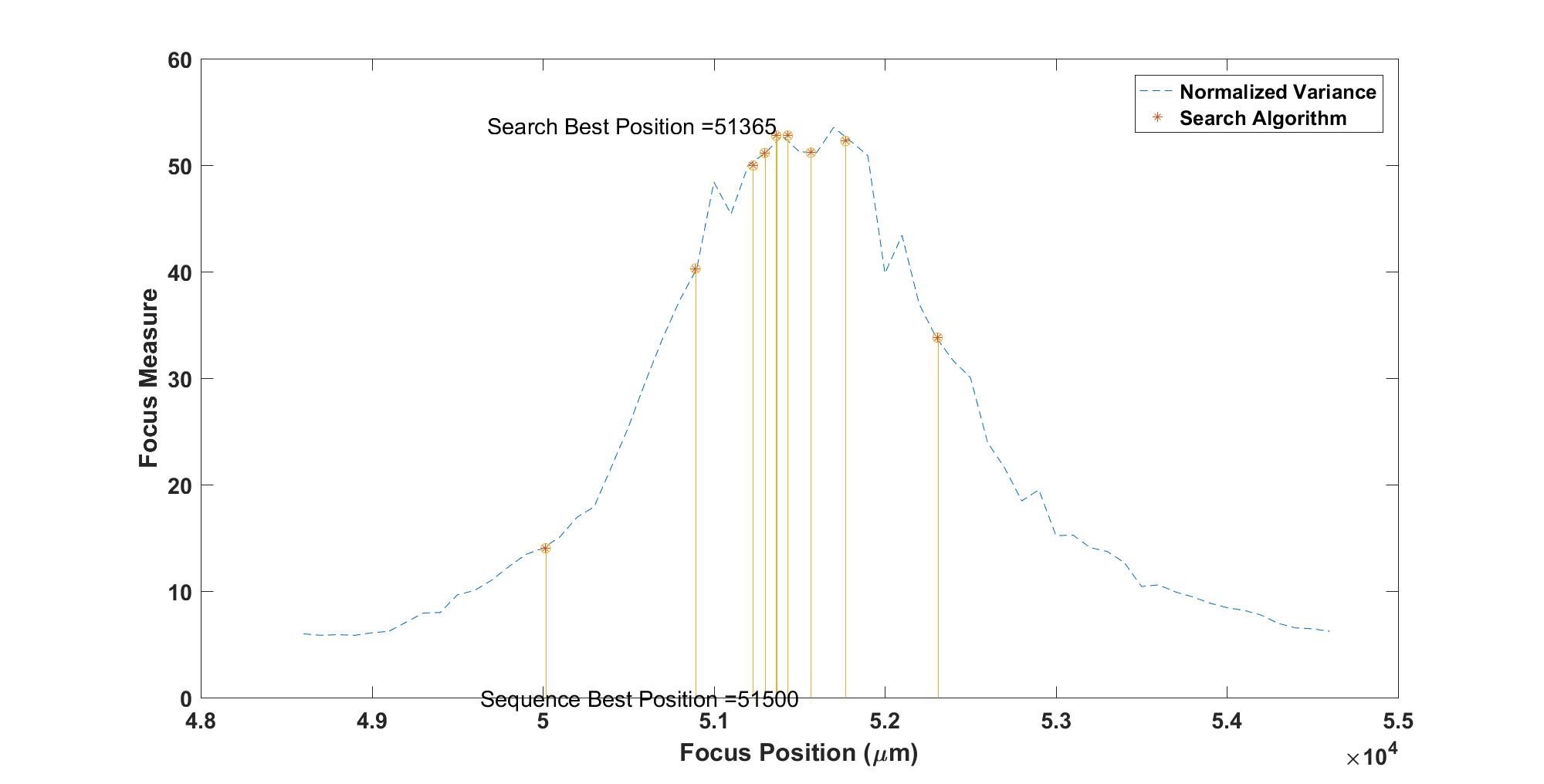}
  \caption{The steps of the Fibonacci search algorithm of the sequence N7788.}
  \label{fig7b}
\end{subfigure}
\vspace*{8pt}
\begin{subfigure}{.5\textwidth}
  \centering
  \includegraphics[width=8cm,height=5.5cm]{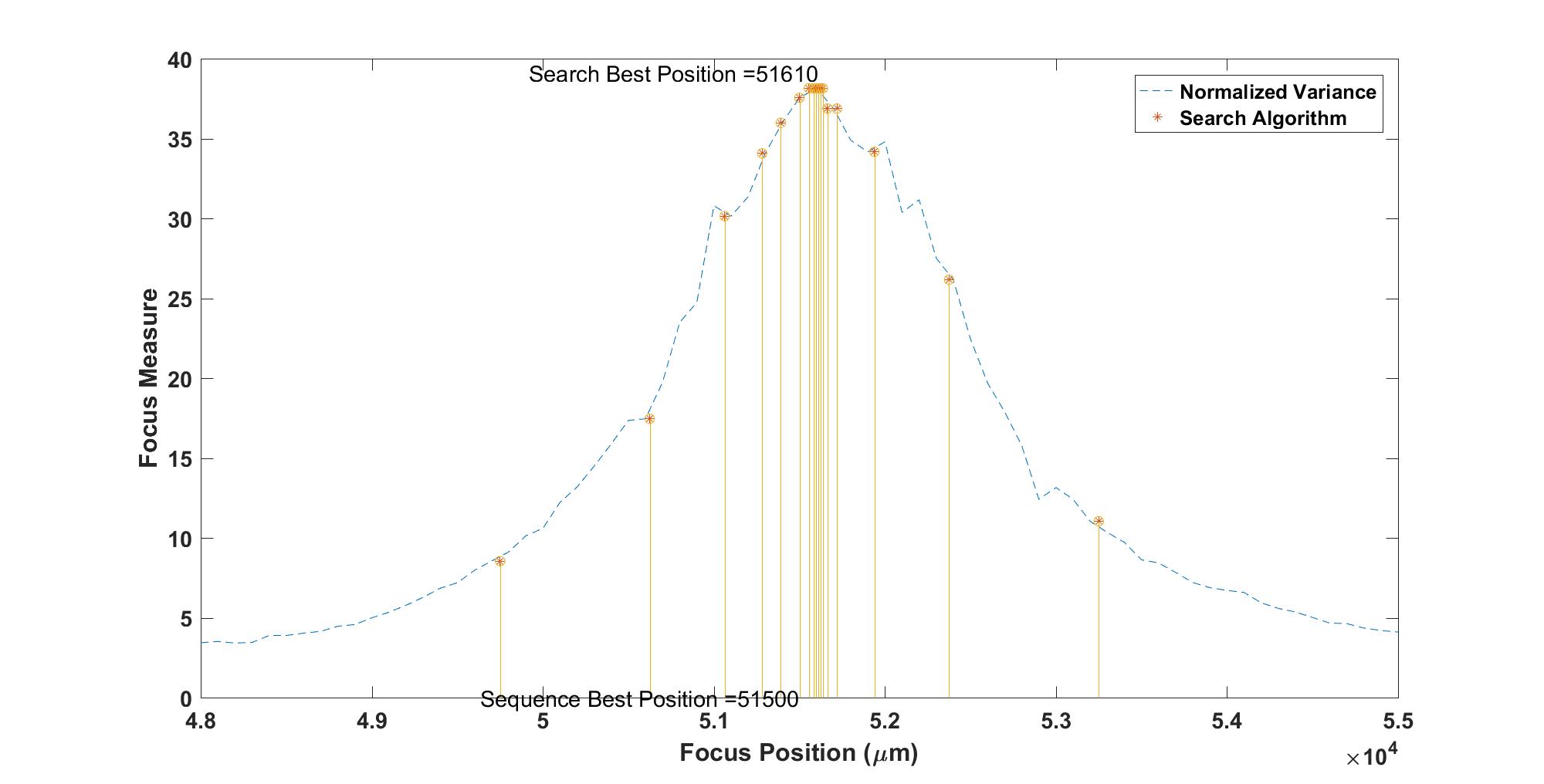}
  \caption{The steps of the Binary search algorithm of the sequence N7789.}
  \label{fig8a}
\end{subfigure}%
\hspace*{8pt}
\begin{subfigure}{.5\textwidth}
  \centering
  \includegraphics[width=8cm,height=5.5cm]{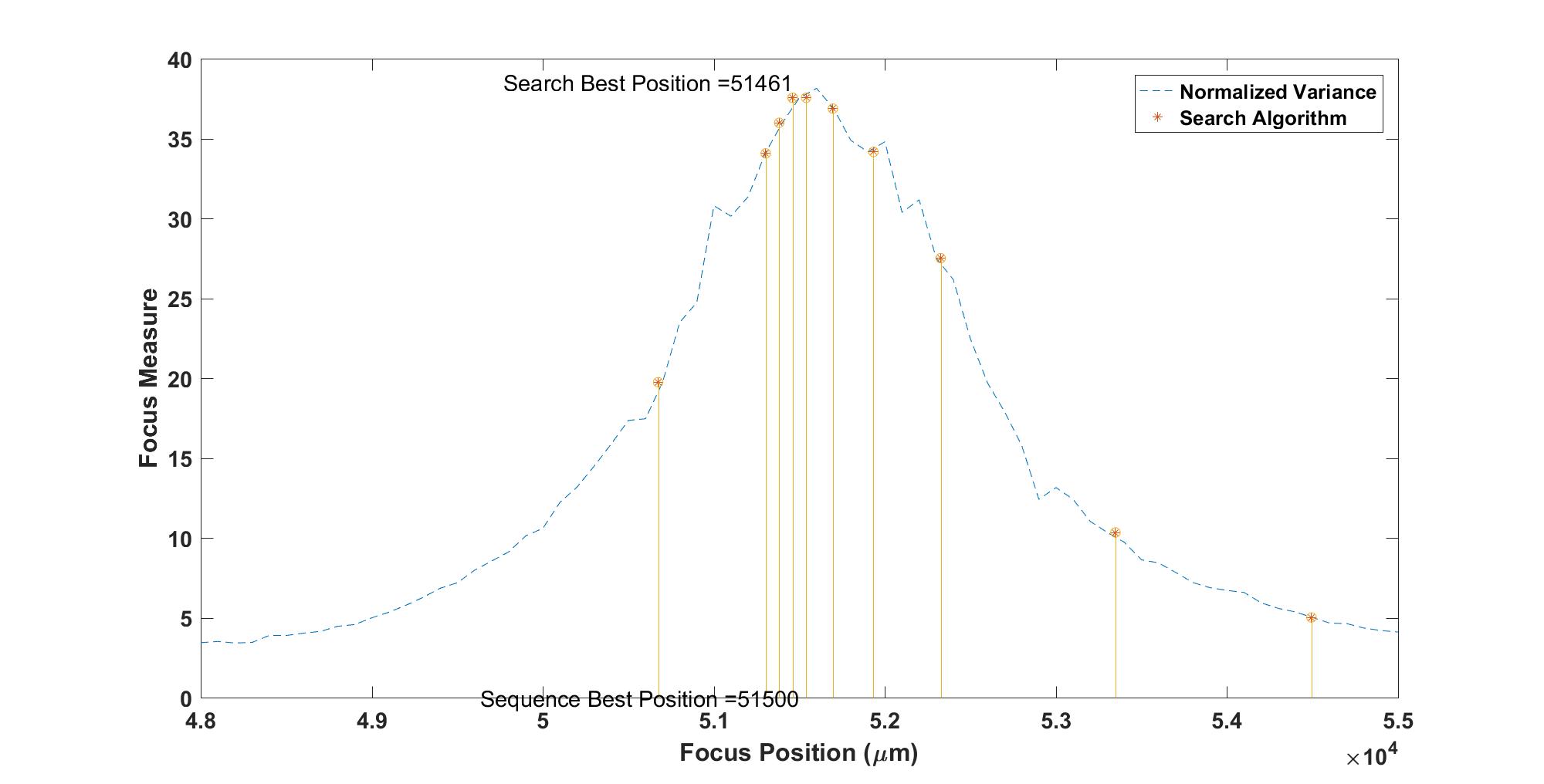}
  \caption{The steps of the Fibonacci search algorithm of the sequence N7789.}
  \label{fig8b}
\end{subfigure}

	\caption{The steps of the Binary and Fibonacci search algorithm of the different sequences.  a) : The steps of the Binary search algorithm of the sequence N7788. b) : The steps of the Fibonacci search algorithm of the sequence N7788. c) : The steps of the Binary search algorithm of the sequence N7789. d) : The steps of the Fibonacci search algorithm of the sequence N7789.}\label{fig8}
\end{figure}

\begin{figure}[H]
\centering
\begin{subfigure}{.5\textwidth}
  \centering
  \includegraphics[width=8cm,height=5.5cm]{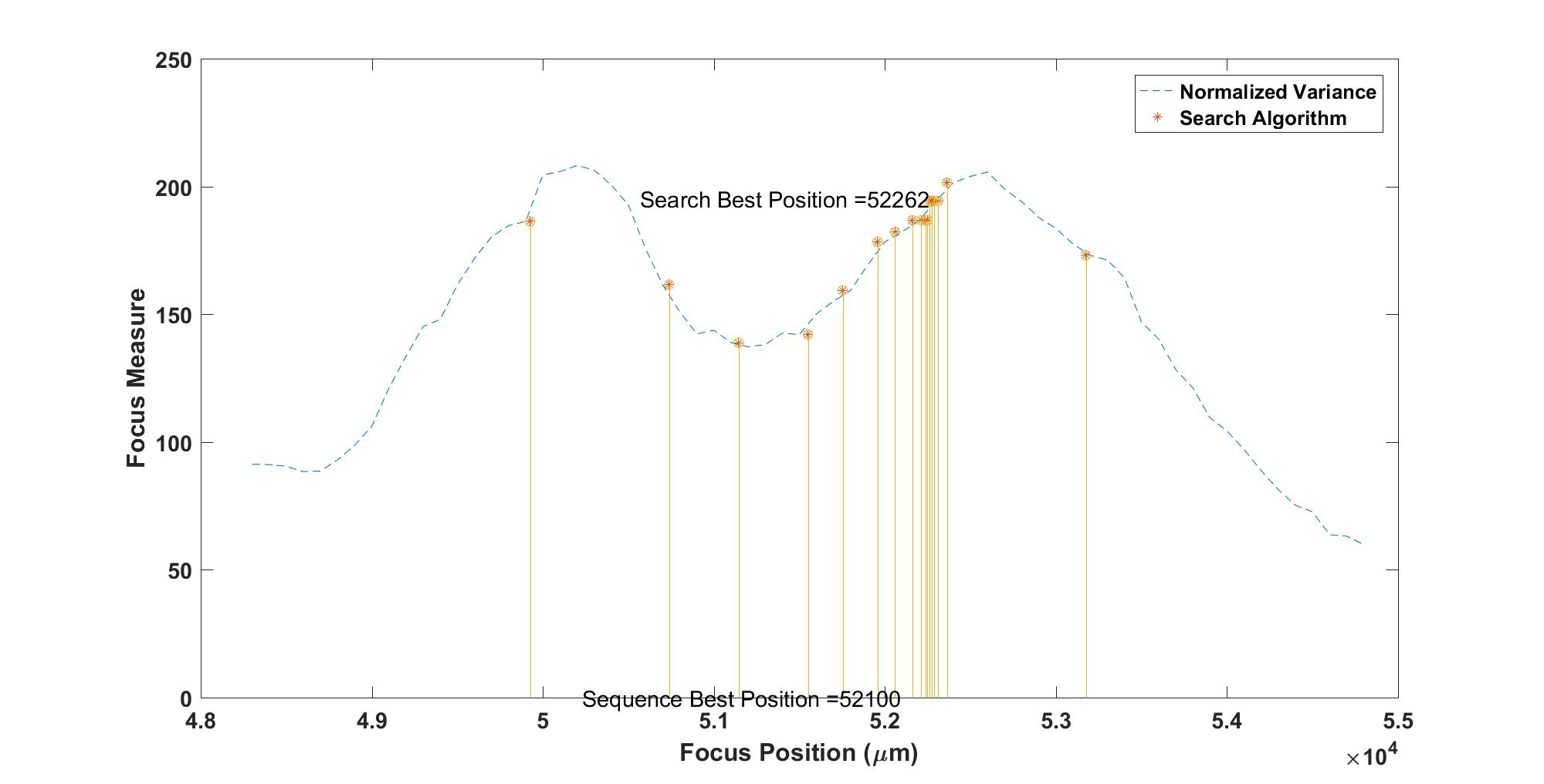}
  \caption{The steps of the Binary search algorithm of the sequence M103.}
  \label{fig4a}
\end{subfigure}%
\hspace*{8pt}
\begin{subfigure}{.5\textwidth}
  \centering
  \includegraphics[width=8cm,height=5.5cm]{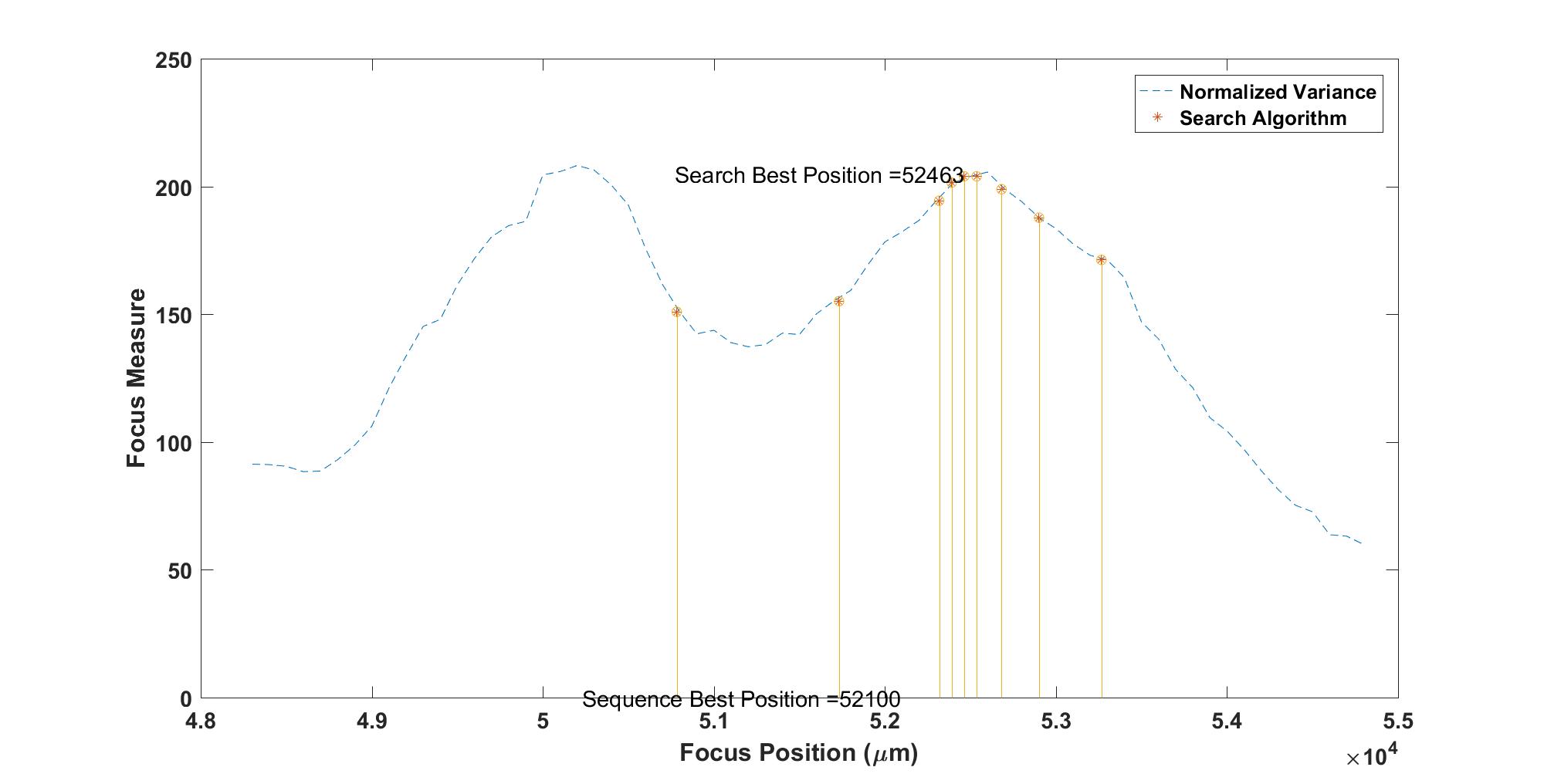}
  \caption{The steps of the Fibonacci search algorithm of the sequence M103.}
  \label{fig4b}
\end{subfigure}

\vspace*{8pt}
\begin{subfigure}{.5\textwidth}
  \centering
  \includegraphics[width=8cm,height=5.5cm]{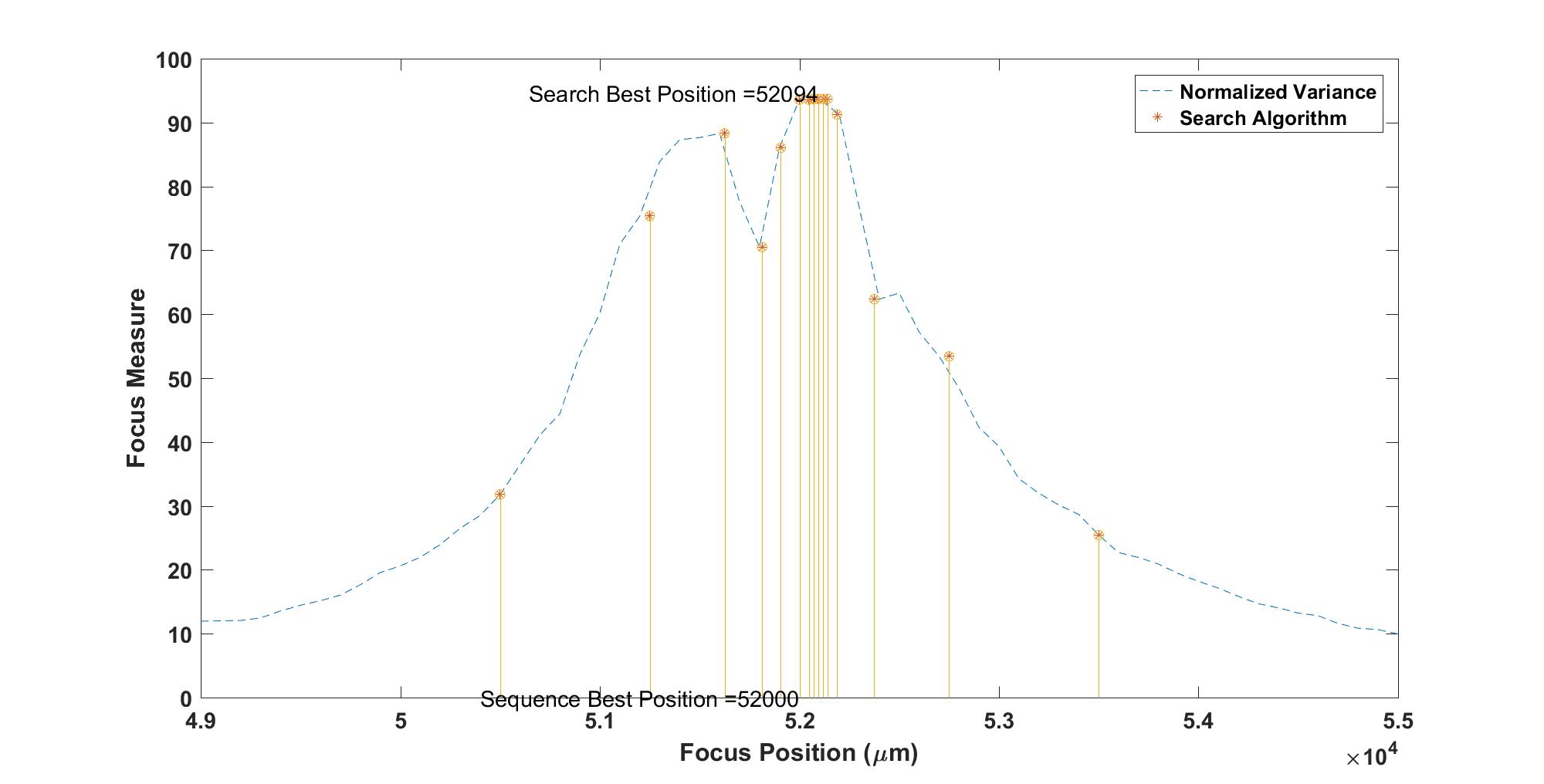}
  \caption{The steps of the Binary search algorithm of the sequence N6793.}
  \label{fig5a}
\end{subfigure}%
\hspace*{8pt}
\begin{subfigure}{.5\textwidth}
  \centering
  \includegraphics[width=8cm,height=5.5cm]{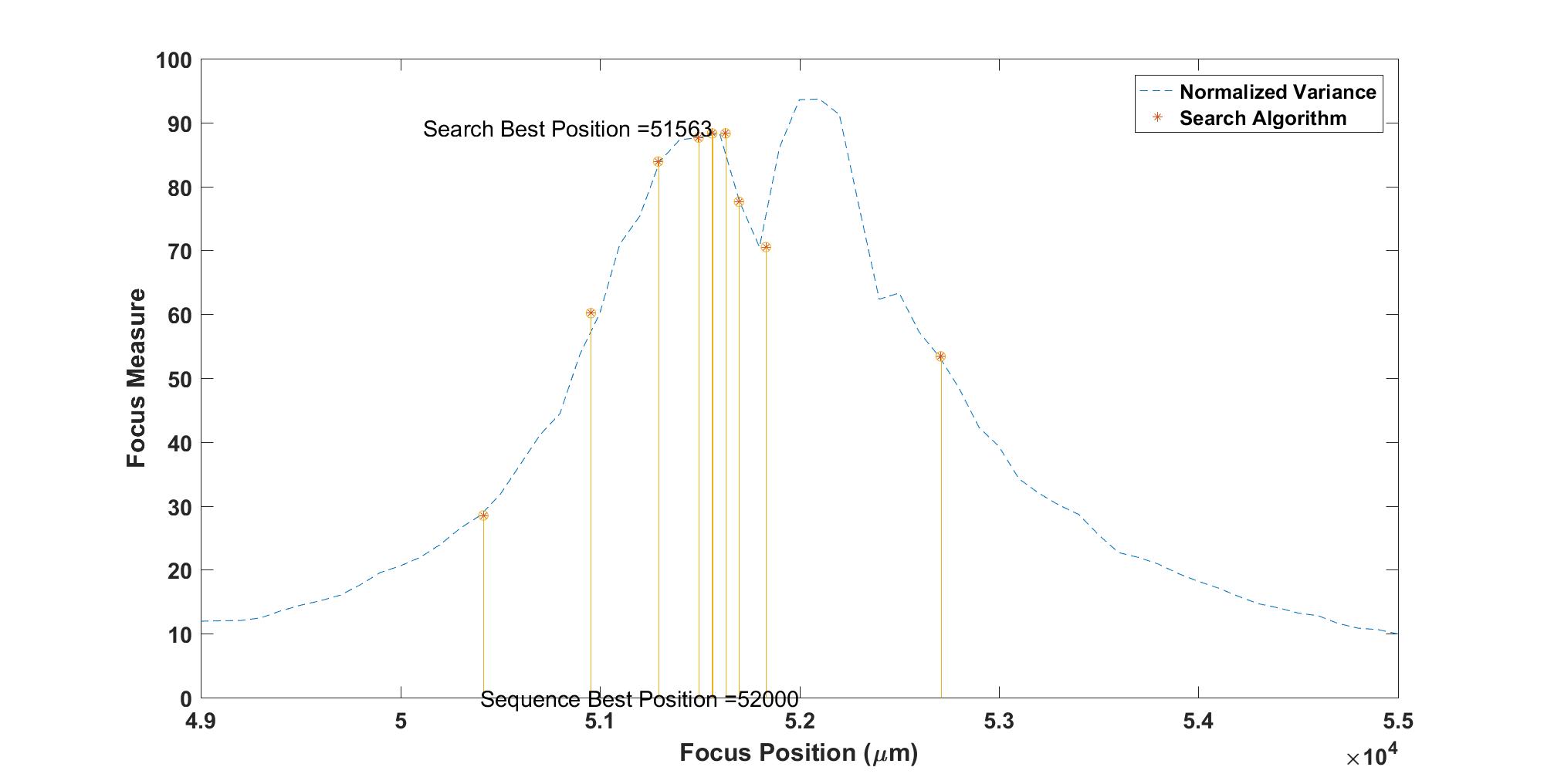}
  \caption{The steps of the Fibonacci search algorithm of the sequence N6793.}
  \label{fig5b}
\end{subfigure}

\vspace*{8pt}
\begin{subfigure}{.5\textwidth}
  \centering
  \includegraphics[width=8cm,height=5.5cm]{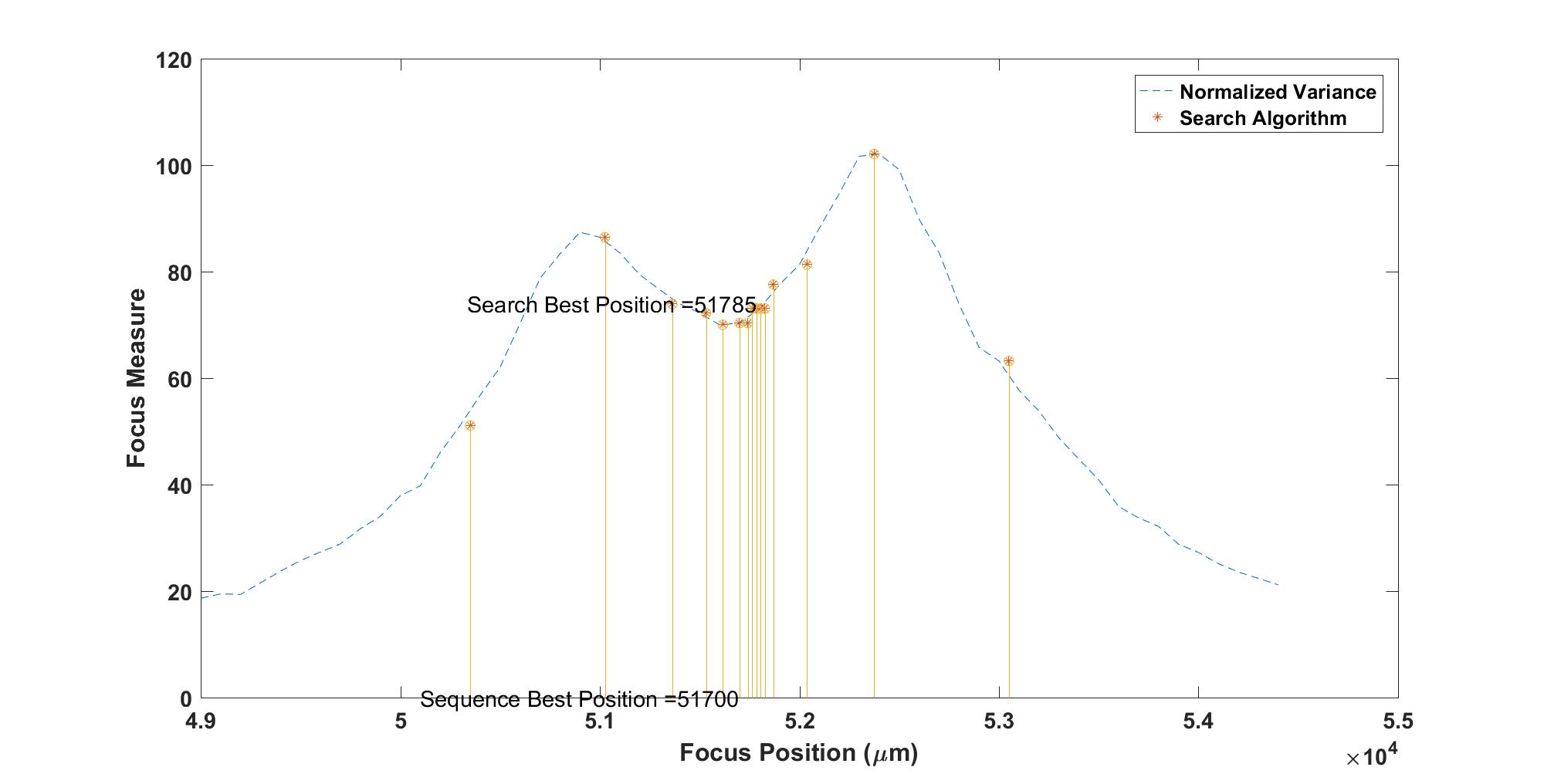}
  \caption{The steps of the Binary search algorithm of the sequence N7067.}
  \label{fig6a}
\end{subfigure}%
\hspace*{8pt}
\begin{subfigure}{.5\textwidth}
  \centering
  \includegraphics[width=8cm,height=5.5cm]{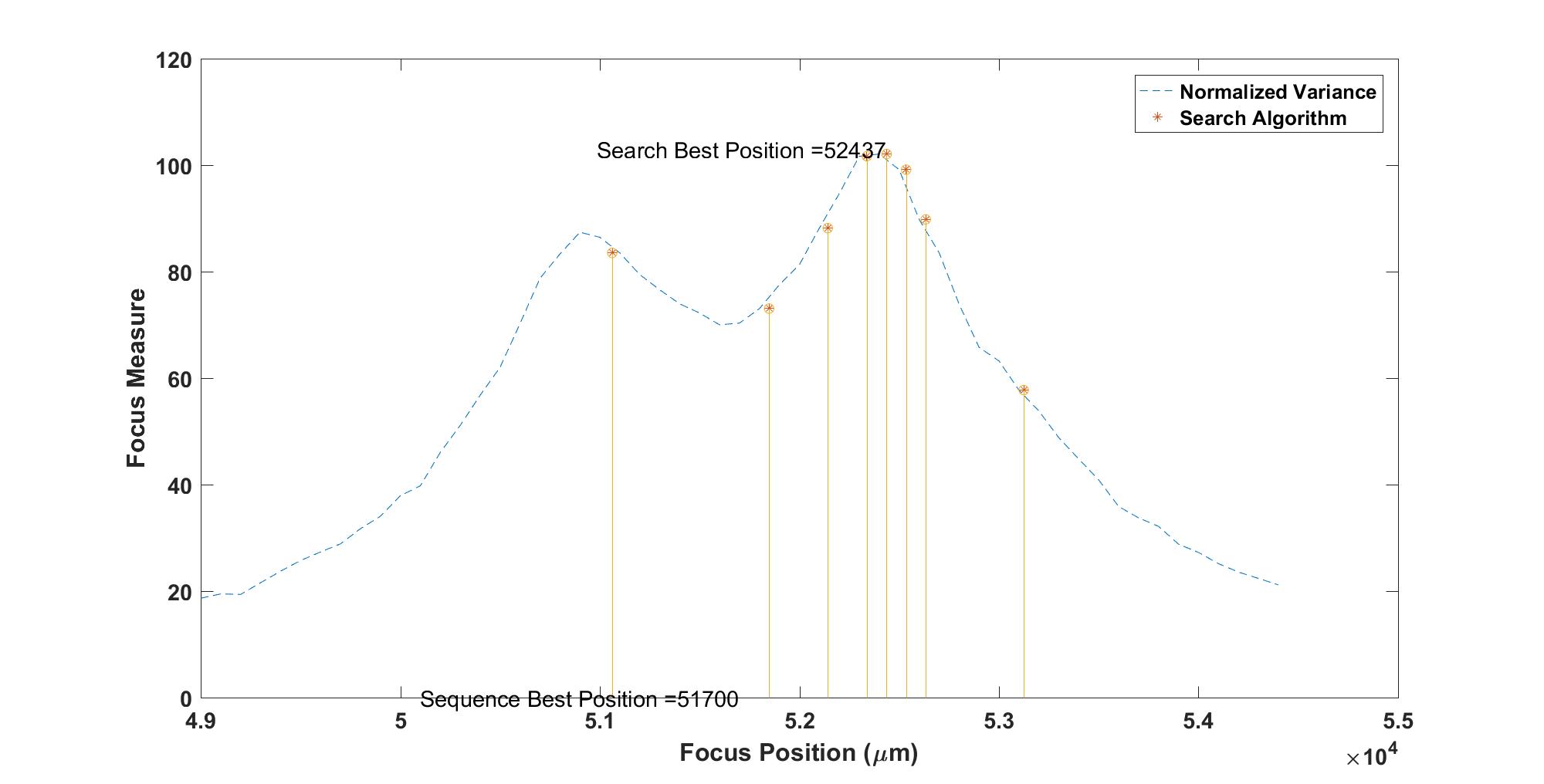}
  \caption{The steps of the Fibonacci search algorithm of the sequence N7067.}
  \label{fig6b}
\end{subfigure}

\vspace*{8pt}

	\caption{The steps of the Binary and Fibonacci search algorithm of the different sequences. (a) : The steps of the Binary search algorithm of the sequence M103. (b) : The steps of the Fibonacci search algorithm of the sequence M103. (c) : The steps of the Binary search algorithm of the sequence N6793. d) : The steps of the Fibonacci search algorithm of the sequence N6793. e) : The steps of the Binary search algorithm of the sequence N7067. f) : The steps of the Fibonacci search algorithm of the sequence N7067. g) : The steps of the Binary search algorithm of the sequence N7788. h) : The steps of the Fibonacci search algorithm of the sequence N7788. i) : The steps of the Binary search algorithm of the sequence N7789. j) : The steps of the Fibonacci search algorithm of the sequence N7789.}\label{fig7}
\end{figure}

Figure \ref{fig7a}, \ref{fig7b}, \ref{fig8a}, \ref{fig8b}, \ref{fig4a}, \ref{fig4b}, \ref{fig5a}, \ref{fig5b}, \ref{fig6a}, and \ref{fig6b}, , and  displays the search steps of the Binary and Fibonacci of the sequence N7788, N7789, M103, N6793, and N7067, respectively. Figure \ref{fig8} clarifies it during the unimodal, while figure \ref{fig7} presents the performance of both algorithms during the bimodal focus measure. Figure \ref{fig4b}, \ref{fig5b}, and \ref{fig6b} shows that the Fibonacci search algorithm is dependent on the performance of the focus measure. In other words, the Fibonacci search required fewer steps than the Binary to reach its final position. However, it is significantly affected by the guide (focus measure) during the search algorithm. \par

\textbf{Thus, we can conclude the Binary search is more effective in the case of large search interval and non-unimodal focus measure. However, the Fibonacci is the best during the short search interval and unimodal focus measure. Consequently, we recommend the Binary search for the technical time to help in creating previous knowledge about the best focus position for different objects. And, in the science runs, the search starts from a position based on previous knowledge, then using the Fibonacci search algorithm with a proper focus measure. }\par

\section{Conclusion}\label{sec5}
The astronomical observations quality is focus-dependent. The search algorithm is important for imaging systems of numerous focus positions. This paper provides a performance evaluation of several search algorithms. The algorithms are applied to five sequences of astronomical images (star-cluster). The sequence was acquired by using the 74-inches telescope of KAO. It includes in-focus and out-of-focus images. The results show that the Binary search is the optimal one. Thus, it is recommended to use the Binary algorithm in searching for the best focus position.


\begin{thebibliography}{9}

\bibitem[{S. Rao.}(2009)]{S. Rao.2009} S. Rao," Engineering Optimization: Theory and Practice", 4th Edition, John Wiley $\&$ Sons, INC., 2009.
\bibitem[{J. He.(2003)}]{J. He.2003} J. He, R. Zhou, Z. Hong, “Modified Fast Climbing Search Auto-Focus Algorithm with Adaptive Step Size Searching
    Technique for Digital Camera”, IEEE Transactions On Consumer Electronics, Vol. 49, No. 2, 2003, PP. 257-262.
\bibitem[{C. Batten.(2000)}]{C. Batten.2000} C. Batten, “Autofocusing and Astigmatism Correction in the Scanning Electron Microscope”, Mphil Thesis, University of Cambridge, 2000. 
\bibitem[{Y. Yao.(2006)}]{Y. Yao.2006} Y. Yao, B. Abidi, N. Doggaz, M. Abidi, “Evaluation of Sharpness Measures and Search Algorithms for the Auto-Focusing of High Magnification Images”, Proc. of SPIE 6246 , 2006.
\bibitem[{M. Subbarao.(1998)}]{M. Subbarao.1998} M. Subbarao, J. Tyan, “Selecting the Optimal Focus Measure for Autofocusing and Depth-From-Focus”, IEEE Transactions
    on Pattern Analysis and Machine Intelligence, Vol. 20, No. 8, 1998, PP. 864-870.
\bibitem[{I. Helmy.(2020)}]{I. Helmy.2020} I. Helmy, F. Elnagahy, A. Hamdy, “Focus Measures Assessment for Astronomical Images”, International Conference on Innovative Trends in Communication and Computer
Engineering (ITCE), 2020.
\bibitem[{Y. Azzam.(2008)}]{Y. Azzam.2008} Y. Azzam, G. Ali, F. Elnagahy, et al., “Current and Future Capabilities of the 74-Inch Telescope of Kottamia Astronomical Observatory in Egypt”, NRIAG Journal of Astronomy and Astrophysics, Special Issue 2008; 271-285.

\end{thebibliography}
\end{document}